\PassOptionsToPackage{hyphens}{url}
\documentclass[superscriptaddress, nofootinbib,  amsmath, amssymb, preprint, a4paper]{revtex4}

\usepackage[margin=1in]{geometry} 
\usepackage[english]{babel}
\usepackage[utf8]{inputenc}
\usepackage[]{graphicx}
\usepackage{xspace}
\usepackage{siunitx}
\usepackage{mhchem}
\DeclareSIUnit\angstrom{\text {Å}}
\usepackage{orcidlink}
\usepackage{hyperref}
\usepackage{fontawesome}
\usepackage{natmove}
\usepackage{placeins}

\usepackage{xparse}
\usepackage{hyperref}
\usepackage{pdfpages}

\newcommand{\githublink}[2]{
  \href{https://github.com/#1/#2}{\faGithub\ \url{#1/#2}}
}

\newcommand{\twitterlink}[1]{
  \href{#1}{\faTwitter}
}

\newcommand{\zenodolink}[1]{
  \href{https://doi.org/#1}{\faArchive\ \url{#1}}
}

\newcommand{\hflogo}{%

\includegraphics[height=.9em]{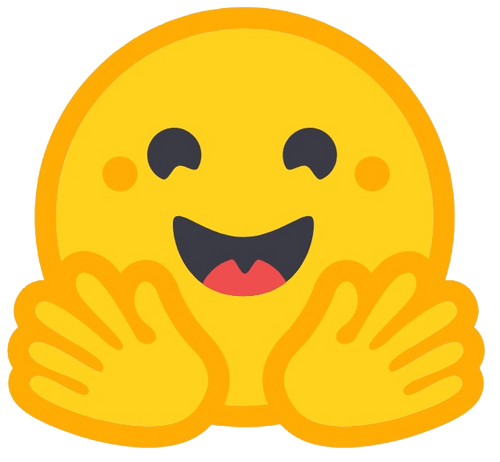}
}

\newcommand{\huggingfacelink}[2]{
  \href{https://huggingface.co/spaces/#1/#2}{\hflogo \url{#1/#2}}
}

\newcommand{\huggingfacehublink}[2]{
  \href{https://huggingface.co/#1/#2}{\hflogo \url{#1/#2}}
}

\hypersetup{hidelinks,colorlinks=true,breaklinks=true,urlcolor=blue,allcolors=blue,pdftitle={Title},pdfauthor={Author}}

\usepackage{listings, xcolor}
\definecolor{codegreen}{rgb}{0,0.6,0}
\definecolor{codegray}{rgb}{0.5,0.5,0.5}
\definecolor{codepurple}{rgb}{0.58,0,0.82}
\definecolor{tqblue}{HTML}{08293d}
\definecolor{backcolour}{HTML}{fefdf5}

\lstdefinestyle{pythonstyle}{
    backgroundcolor=\color{backcolour},   
    commentstyle=\color{codegreen},
    keywordstyle=\color{magenta},
    numberstyle=\tiny\color{codegray},
    stringstyle=\color{codepurple},
    basicstyle=\ttfamily\footnotesize\color{tqblue},
    breakatwhitespace=false,         
    breaklines=true,
    postbreak=\mbox{\textcolor{magenta}{$\hookrightarrow$}\space},                 
    captionpos=b,                    
    keepspaces=true,                 
    numbers=left,                    
    numbersep=5pt,                  
    showspaces=false,                
    showstringspaces=false,
    showtabs=false,                  
    tabsize=2
}

\usepackage{booktabs}
\newcolumntype{C}{>{$}c<{$}}

\AtBeginDocument{%
  \heavyrulewidth=.08em
  \lightrulewidth=.05em
  \cmidrulewidth=.03em
  \belowrulesep=.65ex
  \belowbottomsep=0pt
  \aboverulesep=.4ex
  \abovetopsep=0pt
  \cmidrulesep=\doublerulesep
  \cmidrulekern=.5em
  \defaultaddspace=.5em
}

\usepackage[most]{tcolorbox}

\tcbset {
  base/.style={
    arc=0mm, 
    bottomtitle=0.5mm,
    boxrule=0mm,
    colbacktitle=black!10!white, 
    coltitle=black, 
    fonttitle=\bfseries, 
    left=2.5mm,
    leftrule=1mm,
    right=8.5mm,
    title={#1},
    toptitle=0.75mm, 
    width=\textwidth,
    breakable
  }
}

\definecolor{brandblue}{rgb}{0, 0.27843137254902, 0.466666666666667}
\newtcolorbox{agentinteraction}[1]{
  colframe=brandblue, 
  base={#1}
}

\definecolor{brandbred}{rgb}{0.63921568627451, 0, 0}
\newtcolorbox{agentinteraction2}[1]{
  colframe=brandbred, 
  base={#1}
}

\newtcolorbox{subbox}[1]{
  colframe=black!30!white,
  base={#1}
  }

\usepackage{csquotes}
\usepackage[acronym, nonumberlist]{glossaries}
\makeglossaries

\newacronym{llm}{LLM}{large language model}
\newacronym{gpt}{GPT}{generative pretrained transformer}
\newacronym{api}{API}{application programming interface}
\newacronym{ml}{ML}{machine learning}
\newacronym{lift}{LIFT}{language-interfaced fine-tuning}
\newacronym{icl}{ICL}{in-context learning}
\newacronym{peft}{PEFT}{parameter efficient fine-tuning}
\newacronym{lora}{LoRA}{low-rank adaptors}
\newacronym{gpr}{GPR}{Gaussian process regression}
\newacronym{ga}{GA}{genetic algorithm}
\newacronym{svm}{SVM}{support vector machine}
\newacronym{rf}{RF}{random forest}

\newacronym{ord}{ORD}{Open Reaction Database}
\newacronym{bo}{BO}{Bayesian optimization}
\newacronym{id}{ID}{inverse design}
\newacronym{mad}{MAD}{median absolute deviation}
\newacronym{eln}{ELN}{electronic lab notebook}
\newacronym[shortplural=LIMS]{lims}{LIMS}{laboratory information system}
\newacronym{ui}{UI}{user interface}
\newacronym{nlm}{NLM}{national library of medicine} 
\newacronym{dft}{DFT}{density functional theory}
\newacronym{cot}{COT}{chain of thought}
\newacronym{gui}{GUI}{graphical user interface}
\newacronym{pdb}{PDB}{protein data bank}
\newacronym{rlhf}{RLHF}{reinforcement learning from human feedback}
\newacronym{json}{JSON}{JavaScript object notation}
\newacronym{smiles}{SMILES}{simplified molecular-input line-entry system}
\newacronym{selfies}{SELFIES}{self-referencing embedded strings}
\newacronym{ai}{AI}{artificial intelligence}
\newacronym{nlp}{NLP}{natural language processing}
\newacronym{ner}{NER}{named entity recognition}
\newacronym{cas}{CAS}{Chemical Abstract Services}
\newacronym{mae}{MAE}{mean absolute error}
\newacronym{inchi}{InChI}{international chemical identifier}
\newacronym{mapi}{MAPI}{Materials Project \gls{api}}
\newacronym{rouge}{ROUGE}{Recall-Oriented Understudy for Gisting Evaluation}
\newacronym{html}{HTML}{HyperText Markup Language}
\newacronym{doi}{DOI}{digital object identifier}
\newacronym{ocr}{OCR}{optical character recognition}

\usepackage{tabularx}
\usepackage{cleveref}

\let\originalcite\cite
\renewcommand{\cite}[1]{\unskip~\originalcite{#1}}

\hypersetup{
  colorlinks=false}

\usepackage{setspace}

\usepackage{titlesec}

\titlespacing{\subsection}
    {0pt}{9pt}{6pt}

\usepackage{array}
\usepackage{ragged2e}
\usepackage{nicefrac}
\usepackage[caption=false]{subfig}

\newcolumntype{P}[1]{>{\raggedright\arraybackslash}p{#1}}

\begin{document}

\title{14 Examples of How LLMs Can Transform Materials Science and Chemistry: A Reflection on a Large Language Model Hackathon}


\newcommand{\equalcont}{\thanks{These authors contributed equally}}

\author{Kevin~Maik~Jablonka~\orcidlink{0000-0003-4894-4660}}
\email{mail@kjablonka.com}
\affiliation{Laboratory of Molecular Simulation (LSMO), Institut des Sciences et Ing\'{e}nierie Chimiques, Ecole Polytechnique F\'{e}d\'{e}rale de Lausanne (EPFL), Sion, Valais, Switzerland.}

\author{Qianxiang Ai~\orcidlink{0000-0002-5487-2539}}
\equalcont
\affiliation{Department of Chemical Engineering, Massachusetts Institute of Technology, Cambridge, Massachusetts 02139, United States.}


\author{Alexander~Al-Feghali~\orcidlink{0009-0004-8377-7049}} 
\equalcont
\affiliation{Department of Chemistry, McGill University, Montreal, Quebec, Canada.}

\author{Shruti~Badhwar~\orcidlink{0000-0002-3167-5348}}
\equalcont
\affiliation{Reincarnate Inc.}  

\author{Joshua~D.~Bocarsly~\orcidlink{0000-0002-7523-152X}}
\equalcont
 \affiliation{Yusuf Hamied Department of Chemistry, University of Cambridge, Lensfield Road, Cambridge, CB2 1EW, United Kingdom.} 
 

\author{Andres~M~Bran~\orcidlink{0000-0002-4432-3667}}
\equalcont
\affiliation{Laboratory of Artificial Chemical Intelligence (LIAC), Institut des Sciences et Ing\'{e}nierie Chimiques, Ecole Polytechnique F\'{e}d\'{e}rale de Lausanne (EPFL), Lausanne, Switzerland.} 
\affiliation{National Centre of Competence in Research (NCCR) Catalysis, Ecole Polytechnique F\'{e}d\'{e}rale de Lausanne (EPFL), Lausanne, Switzerland.}

\author{Stefan~Bringuier~\orcidlink{0000-0001-6753-1437}}
\equalcont
\affiliation{Independent Researcher, San Diego, CA, United States.}

\author{L.~Catherine~Brinson~\orcidlink{0000-0003-2551-1563}}
\equalcont
\affiliation{Mechanical Engineering and Materials Science, Duke University, United States.}

\author{Kamal~Choudhary~\orcidlink{0000-0001-9737-8074}}
\equalcont
 \affiliation{ 
Material Measurement Laboratory, National Institute of Standards and Technology, Maryland, 20899, United States.
}%

\author{Defne~Circi~\orcidlink{0000-0002-5761-0198}}
\equalcont
\affiliation{Mechanical Engineering and Materials Science, Duke University, United States.}

\author{Sam~Cox~\orcidlink{0000-0002-4441-9327}}
\equalcont
\affiliation{Department of Chemical Engineering, University of Rochester, United States.}

\author{Wibe~A.~de~Jong~\orcidlink{0000-0002-7114-8315}} 
\equalcont
\affiliation{Applied Mathematics and Computational Research Division, Lawrence Berkeley National Laboratory, Berkeley, CA 94720, United States.
}

\author{Matthew~L.~Evans~\orcidlink{0000-0002-1182-9098}}
\equalcont
 \affiliation{Institut de la Matière Condensée et des Nanosciences (IMCN), UCLouvain, Chemin des Étoiles 8, Louvain-la-Neuve, 1348, Belgium.}
 \affiliation{Matgenix SRL, 185 Rue Armand Bury, 6534 Gozée, Belgium.}

 \author{Nicolas~Gastellu~\orcidlink{0000-0002-4052-076X}}
 \equalcont
\affiliation{Department of Chemistry, McGill University, Montreal, Quebec, Canada.}

\author{Jerome~Genzling~\orcidlink{0009-0007-4728-1478}} 
\equalcont
\affiliation{Department of Chemistry, McGill University, Montreal, Quebec, Canada.}

\author{Mar\'ia~Victoria~Gil~\orcidlink{0000-0002-2258-3011}}
\equalcont
\affiliation{Instituto de Ciencia y Tecnolog\'ia del Carbono (INCAR), CSIC, Francisco Pintado Fe 26, 33011 Oviedo, Spain.}

\author{Ankur~K.~Gupta~\orcidlink{0000-0002-3128-9535}}
\equalcont
\affiliation{Applied Mathematics and Computational Research Division, Lawrence Berkeley National Laboratory, Berkeley, CA 94720, United States.
}

\author{Zhi~Hong~\orcidlink{0000-0002-5250-1347}}
\equalcont
\affiliation{Department of Computer Science, University of Chicago, Chicago, Illinois 60637, United States.}

\author{Alishba~Imran}
\equalcont
\affiliation{Computer Science, University of California, Berkeley, Berkeley CA 94704, United States.}

\author{Sabine~Kruschwitz~\orcidlink{0000-0002-6296-4417}}
\equalcont
\affiliation{Bundesanstalt für Materialforschung und -prüfung, Unter den Eichen 87, 12205 Berlin, Germany.}

\author{Anne~Labarre~\orcidlink{0000-0003-4939-3928}}
\equalcont
\affiliation{Department of Chemistry, McGill University, Montreal, Quebec, Canada.}

\author{Jakub~Lála~\orcidlink{0000-0002-5424-5260}}
\equalcont
\affiliation{Francis Crick Institute, 1 Midland Rd, London NW1 1AT, United Kingdom.}
 
\author{Tao~Liu~\orcidlink{0000-0002-1082-5570}}
\equalcont
\affiliation{Department of Chemistry, McGill University, Montreal, Quebec, Canada.}

\author{Steven~Ma~\orcidlink{0000-0006-9448-7332}}
\equalcont
\affiliation{Department of Chemistry, McGill University, Montreal, Quebec, Canada.}

\author{Sauradeep~Majumdar~\orcidlink{0000-0002-2095-3082}}
\equalcont
\affiliation{Laboratory of Molecular Simulation (LSMO), Institut des Sciences et Ing\'{e}nierie Chimiques, Ecole Polytechnique F\'{e}d\'{e}rale de Lausanne (EPFL), Sion, Valais, Switzerland.}

\author{Garrett~W.~Merz~\orcidlink{0000-0003-4737-3931}}
\equalcont
\affiliation{American Family Insurance Data Science Institute, University of Wisconsin-Madison, Madison WI 53706, United States.}

\author{Nicolas~Moitessier~\orcidlink{0000-0001-6933-2079}}
\equalcont
\affiliation{Department of Chemistry, McGill University, Montreal, Quebec, Canada.}

\author{Elias~Moubarak~\orcidlink{0000-0001-8271-6800}}
\equalcont
\affiliation{Laboratory of Molecular Simulation (LSMO), Institut des Sciences et Ing\'{e}nierie Chimiques, Ecole Polytechnique F\'{e}d\'{e}rale de Lausanne (EPFL), Sion, Valais, Switzerland.}

\author{Beatriz~Mouriño~\orcidlink{0000-0003-1670-3985}}
\equalcont
\affiliation{Laboratory of Molecular Simulation (LSMO), Institut des Sciences et Ing\'{e}nierie Chimiques, Ecole Polytechnique F\'{e}d\'{e}rale de Lausanne (EPFL), Sion, Valais, Switzerland.}

\author{Brenden~Pelkie~\orcidlink{0000-0001-7638-6366}}
\equalcont
\affiliation{Department of Chemical Engineering, University of Washington, Seattle, WA 98105, United States.} 

\author{Michael~Pieler~\orcidlink{0000-0001-9186-7045}}
\equalcont
\affiliation{OpenBioML.org}
\affiliation{Stability.AI}

\author{Mayk~Caldas~Ramos~\orcidlink{0000-0001-5336-2847}}
\equalcont
\affiliation{Department of Chemical Engineering, University of Rochester, United States.}

\author{Bojana~Ranković~\orcidlink{0000-0002-1476-6686}}
\equalcont
\affiliation{Laboratory of Artificial Chemical Intelligence (LIAC), Institut des Sciences et Ing\'{e}nierie Chimiques, Ecole Polytechnique F\'{e}d\'{e}rale de Lausanne (EPFL), Lausanne, Switzerland.}
\affiliation{National Centre of Competence in Research (NCCR) Catalysis, Ecole Polytechnique F\'{e}d\'{e}rale de Lausanne (EPFL), Lausanne, Switzerland.}

\author{Samuel~G.~Rodriques~\orcidlink{0000-0002-2509-0861}}
\equalcont
\affiliation{Francis Crick Institute, 1 Midland Rd, London NW1 1AT, United Kingdom.}

\author{Jacob~N.~Sanders~\orcidlink{0000-0002-2196-4234}}
\equalcont
\affiliation{Department of Chemistry and Biochemistry, University of California, Los Angeles, CA 90095, United States.}



\author{Philippe~Schwaller~\orcidlink{0000-0003-3046-6576}}
\equalcont
\affiliation{Laboratory of Artificial Chemical Intelligence (LIAC), Institut des Sciences et Ing\'{e}nierie Chimiques, Ecole Polytechnique F\'{e}d\'{e}rale de Lausanne (EPFL), Lausanne, Switzerland.}
\affiliation{National Centre of Competence in Research (NCCR) Catalysis, Ecole Polytechnique F\'{e}d\'{e}rale de Lausanne (EPFL), Lausanne, Switzerland.}

\author{Marcus~Schwarting}
\equalcont
 \affiliation{Department of Computer Science, University of Chicago, Chicago IL 60490, United States.}

\author{Jiale~Shi~\orcidlink{0000-0002-5447-3925}}
\equalcont
\affiliation{Department of Chemical Engineering, Massachusetts Institute of Technology, Cambridge, Massachusetts 02139, United States.}
 
 \author{Berend~Smit~\orcidlink{0000-0003-4653-8562}}
\equalcont
\affiliation{Laboratory of Molecular Simulation (LSMO), Institut des Sciences et Ing\'{e}nierie Chimiques, Ecole Polytechnique F\'{e}d\'{e}rale de Lausanne (EPFL), Sion, Valais, Switzerland.}

\author{Ben~E.~Smith~\orcidlink{0000-0001-9673-2449}}
\equalcont
 \affiliation{Yusuf Hamied Department of Chemistry, University of Cambridge, Lensfield Road, Cambridge, CB2 1EW, United Kingdom.} 

\author{Joren~Van~Herck~\orcidlink{009-0005-5108-5061}}
\equalcont
\affiliation{Laboratory of Molecular Simulation (LSMO), Institut des Sciences et Ing\'{e}nierie Chimiques, Ecole Polytechnique F\'{e}d\'{e}rale de Lausanne (EPFL), Sion, Valais, Switzerland.}

%
\author{Christoph~Völker~\orcidlink{0000-0002-0985-0074}} 
\equalcont
\affiliation{Bundesanstalt für Materialforschung und -prüfung, Unter den Eichen 87, 12205 Berlin, Germany.}

\author{Logan~Ward~\orcidlink{0000-0002-1323-5939}}
\equalcont
\affiliation{Data Science and Learning Division, Argonne National Lab, United States.}

\author{Sean~Warren~\orcidlink{0000-0002-3670-0354}}
\equalcont

\author{Benjamin~Weiser~\orcidlink{0000-0002-3770-7224}}
\equalcont
\affiliation{Department of Chemistry, McGill University, Montreal, Quebec, Canada.}

\author{Sylvester~Zhang}
\equalcont
\affiliation{Department of Chemistry, McGill University, Montreal, Quebec, Canada.}

\author{Xiaoqi~Zhang\orcidlink{0000-0002-6507-6490}}
\equalcont
\affiliation{Laboratory of Molecular Simulation (LSMO), Institut des Sciences et Ing\'{e}nierie Chimiques, Ecole Polytechnique F\'{e}d\'{e}rale de Lausanne (EPFL), Sion, Valais, Switzerland.}

\author{Ghezal~Ahmad~Zia~\orcidlink{0000-0002-9082-9423}} 
\equalcont
\affiliation{Bundesanstalt für Materialforschung und -prüfung, Unter den Eichen 87, 12205 Berlin, Germany.}


\author{Aristana~Scourtas~\orcidlink{0000-0002-3917-605X}}
\affiliation{Globus, University of Chicago, Data Science and Learning Division, Argonne National Lab, United States.}

\author{KJ~Schmidt}
\affiliation{Globus, University of Chicago, Data Science and Learning Division, Argonne National Lab, United States.}

\author{Ian~Foster~\orcidlink{0000-0003-2129-5269}}
\affiliation{Department of Computer Science, University of Chicago, Data Science and Learning Division, Argonne National Lab, United States.}

\author{Andrew~D.~White~\orcidlink{0000-0002-6647-3965}}
\affiliation{Department of Chemical Engineering, University of Rochester, United States.}

\author{Ben~Blaiszik~\orcidlink{0000-0002-5326-4902}}
\email{blaiszik@uchicago.edu}
\affiliation{Globus, University of Chicago, Data Science and Learning Division, Argonne National Lab, United States.}

\begin{abstract}
Large-language models (LLMs) such as GPT-4 caught the interest of many scientists.
Recent studies suggested that these models could be useful in chemistry and materials science. To explore these possibilities, we organized a hackathon.

This article chronicles the projects built as part of this hackathon. Participants employed LLMs for various applications, including predicting properties of molecules and materials, designing novel interfaces for tools, extracting knowledge from unstructured data, and developing new educational applications.

The diverse topics and the fact that working prototypes could be generated in less than two days highlight that LLMs will profoundly impact the future of our fields.  
The rich collection of ideas and projects also indicates that the applications of LLMs are not limited to materials science and chemistry but offer potential benefits to a wide range of scientific disciplines.
\end{abstract}

\maketitle


\section{Introduction}
The intersection of \gls{ml} with chemistry and materials science has witnessed remarkable advancements in recent years \cite{Butler_2018, Moosavi_2020, Morgan_2020, ramprasad2017machine, Schmidt_2019,choudhary2022recent, Jablonka_2020,shi2022predicting,Shi2023transfer}. 
Much progress has been made in using \gls{ml} to, e.g., accelerate simulations \cite{No__2020, Batzner_2022} or to directly predict properties or compounds for a given application \cite{sanchez2018inverse}. Thereby, developing custom, hand-crafted models for any given application is still common practice.  
Since science rewards doing novel things for the first time, we now face a deluge of tools and machine-learning models for various tasks. 
These tools commonly require input data in their own \emph{rigid, well-defined form} (e.g., a table with specific columns or images from a specific microscope with specific dimensions). 
Further, they typically also report their outputs in non-standard and sometimes proprietary forms.

This rigidity sharply contrasts the standard practice in the (experimental) molecular and materials sciences, which is intrinsically \emph{fuzzy and highly context-dependent} \cite{Gonthier_2012}.
For instance, researchers have many ways to refer to a molecule (e.g., IUPAC name, conventional name, \gls{smiles} \cite{weininger1988smiles}) and to report results and procedures. In particular, for the latter, it is known that small details such as the order of addition or the strength of stirring (e.g., \enquote{gently} vs.\ \enquote{strongly}) are crucial in determining the outcome of reactions.
We do not have a natural way to deal with this fuzziness, and often a conversion into structured tabular form (the conventional input format for \gls{ml} models) is impossible. 
Our current \enquote{solution} is to write conversion programs and chain many tools with plenty of application-specific \enquote{glue code} to enable scientific workflows. 
However, this fuzziness chemistry and heterogeneity of tools have profound consequences: A never-ending stream of new file formats, interfaces, and interoperability tools exists, and users cannot keep up with learning \cite{Jablonka_2022_eln}. 
In addition, almost any transformation of highly context-dependent text (e.g., description of a reaction procedure) into structured, tabular form will lead to a loss of information.   

One of the aims of this work is to demonstrate how 
\glspl{llm} such as the \gls{gpt}-4 \cite{DBLP:journals/corr/abs-2108-07258, vaswani2017attention, chowdhery2022palm, hoffmann2022training, brown2020language, edwards2022translation}, can be used to address these challenges.  
Foundation models such as \glspl{gpt} are general-purpose technologies \cite{eloundou2023gpts} that can solve tasks they have not explicitly been trained on\cite{srivastava2022beyond, bubeck2023sparks}, use tools \cite{schick2023toolformer, karpas2022mrkl, shen2023hugginggpt}, and be grounded in knowledge bases \cite{paperqa, Liu_LlamaIndex_2022}. 
As we also show in this work, they provide new pathways of exploration, new opportunities for flexible interfaces, and may be used to effectively solve certain tasks themselves; e.g., we envision \glspl{llm} enabling non-experts to program (\enquote{malleable software}) using natural language as the \enquote{programming language} \cite{kaparthy}, extract structured information, and create digital assistants that make our tools interoperable---all based on unstructured, natural-language inputs.

Inspired by early reports on the use of these \glspl{llm} in chemical research \cite{Hocky_2022,Jablonka_2023,white2023assessment, Ramos2023}, we organized a virtual hackathon event focused on understanding the applicability of \glspl{llm} to materials science and chemistry. 
The hackathon aimed to explore the multifaceted applications of \glspl{llm} in materials science and chemistry and encourage creative solutions to some of the pressing challenges in the field.
This article showcases some of the projects (\Cref{tab:projects}) developed during the hackathon. 

One of the conclusions of this work is that without these \glspl{llm}, such projects would take many months. 
The diversity of topics these projects address illustrates the broad applicability of \glspl{llm}; the projects touch many different aspects of materials science and chemistry, from the wet lab to  the computational chemistry lab, software interfaces, and even the classroom.
While the examples below are not yet polished products, the simple observation that such capabilities could be created in hours underlines that we need to start thinking about how \glspl{llm} will impact the future of materials science, chemistry, and beyond \cite{White_2023}. The diverse applications show that \glspl{llm} are here to stay and are likely a foundational capability that will be integrated into most aspects of the research process. Even so, the pace of the developments highlights that we are only beginning to scratch the surface of what \glspl{llm} can do for chemistry and materials science.  

\Cref{tab:projects} lists the different projects created in this collaborative effort across eight countries and 22 institutions (SI section V). 
One might expect that 1.5 days of intense collaborations would, at best, allow a cursory exploration of a topic. However, the diversity of topics and the diversity in the participants' expertise, combined with the need to deliver a working prototype (within a short window of time) and the ease of prototyping with \glspl{llm}, generated not only many questions but also pragmatic solutions. In the remainder of this article, we focus on the insights we obtained from this collective effort. For the details of each project, we refer to the SI. 

\begingroup
\renewcommand{\arraystretch}{1.5}
\begin{table*}
    \tiny
    \caption{\emph{Overview of the developed tools and links to source code repositories.} Full descriptions of the projects can be found in the Supplementary Material.}

    \begin{tabularx}{\linewidth}{P{0.1em}P{18em}P{20em}P{22em}}
        \toprule 
        & name & authors & links \\ 
        \midrule 
        \multicolumn{4}{l}{\emph{Predictive modeling}} \\

         &Accurate Molecular Energy Predictions  & Ankur~K.~Gupta, Garrett~W.~Merz, Alishba~Imran, Wibe~A.~de~Jong & \githublink{ankur56}{ChemLoRA}\newline \zenodolink{10.5281/zenodo.8104930} \\
         & Text2Concrete   & Sabine~Kruschwitz, Christoph~Völker, Ghezal~Ahmad~Zia & \githublink{ghezalahmad}{LLMs-for-the-Design-of-Sustainable-Concretes}\newline \zenodolink{10.5281/zenodo.8091195} \\
        & Molecule Discovery by Context & Zhi~Hong, Logan~Ward 
        &\vspace*{-2em} \huggingfacehublink{globuslabs}{ScholarBERT-XL}\newline \zenodolink{10.5281/zenodo.8122087} \\
         & Genetic algorithm without genes &Benjamin~Weiser, Jerome~Genzling, Nicolas~Gastellu, Sylvester~Zhang, Tao~Liu, Alexander~Al-Feghali, Nicolas~Moitessier, Anne~Labarre, Steven~Ma & \githublink{BenjaminWeiser}{LLM-Guided-GA}\newline \zenodolink{10.5281/zenodo.8125541} \\
         & Text-template paraphrasing & Michael~Pieler & \githublink{micpie}{text-template-paraphrasing-chemistry} \newline \zenodolink{10.5281/zenodo.8093615} \\
        \midrule
        \multicolumn{4}{l}{\emph{Automation and novel interfaces}} \\
        &BOLLaMa & Bojana~Ranković, Andres~M.~Bran, Philippe~Schwaller
  & \githublink{doncamilom}{BOLLaMa}\newline \zenodolink{10.5281/zenodo.8096827} \\ 
  &sMolTalk & Jakub~L\'{a}la, Sean~Warren, Samuel~G.~Rodriques &\githublink{jakublala}{smoltalk-legacy} \newline \zenodolink{10.5281/zenodo.8081749} \\
  &MAPI-LLM &Mayk~Caldas~Ramos, Sam~Cox, Andrew~White & \githublink{maykcaldas}{MAPI_LLM} \newline \vspace*{-1em} \huggingfacelink{maykcaldas}{MAPI_LLM} \newline \zenodolink{10.5281/zenodo.8097336} \\
  & Conversational \gls{eln} interface (\texttt{Whinchat}) & Joshua~D.~Bocarsly, Matthew~L.~Evans and Ben~E.~Smith & \githublink{the-grey-group}{datalab}\newline \zenodolink{10.5281/zenodo.8127782} \\
        \midrule 
        \multicolumn{4}{l}{\emph{Knowledge Extraction}}\\
        & InsightGraph & Defne~Circi, Shruti~Badhwar & \githublink{defnecirci}{InsightGraph} \newline \zenodolink{10.5281/zenodo.8092575} \\
        & Extracting Structured Data from Free-form Organic Synthesis Text & Qianxiang~Ai, Jacob~N.~Sanders, Jiale~Shi, Stefan~Bringuier, Brenden~Pelkie, Marcus~Schwarting & \githublink{qai222}{LLM_organic_synthesis}\newline  \zenodolink{10.5281/zenodo.8091902} \\
        & TableToJson: Structured information from scientific data in tables & Mar\'ia~Victoria~Gil 
        & \githublink{vgvinter}{TableToJson} \newline \zenodolink{10.5281/zenodo.8093731}  \\
        & AbstractToTitle \& TitleToAbstract: text summarization and generation & Kamal~Choudhary 
        & \githublink{usnistgov}{chemnlp} \newline \zenodolink{10.5281/zenodo.8122419} \\
     \midrule 
        \multicolumn{4}{l}{\emph{Education}} \\
        & I-Digest & Beatriz~Mouriño,  Elias~Moubarak, Joren~Van~Herck, Sauradeep~Majumdar, Xiaoqi~Zhang & \githublink{XiaoqZhang}{i-Digest} \newline \zenodolink{10.5281/zenodo.8080962}\\
        \bottomrule
    \end{tabularx}  

    \label{tab:projects}
\end{table*}
\endgroup 

We have grouped the projects into four categories: \emph{1. predictive modeling}, \emph{2. automation and novel interfaces}, \emph{3. knowledge extraction}, and \emph{4. education}. The projects in the \emph{predictive modeling} category use \glspl{llm} for classification and regression tasks---and also investigate ways to incorporate established concepts such as $\Delta$-ML\cite{ramakrishnan2015big} or novel concepts such as \enquote{fuzzy} context into the modeling. The \emph{automation and novel interfaces} projects show that natural language might be the universal \enquote{glue} connecting our tools---perhaps in the future, we will need not to focus on new formats or standards but rather use natural language descriptions to connect across the existing diversity and different modalities \cite{White_2023}.

\Glspl{llm} can also help make knowledge more accessible, as the projects in the \enquote{knowledge extraction} category show; they can extract structured information from unstructured text. 
In addition, as the project in the \enquote{education} category shows, \glspl{llm} can also offer new educational opportunities.

\subsection{Predictive modeling}
Predictive modeling is a common application of \gls{ml} in chemistry. 
Based on the \gls{lift} framework \cite{Dinh2022-oj}, \citet{Jablonka_2023} have shown that \glspl{llm} can be employed to predict various chemical properties, such as solubility or HOMO-LUMO gaps based on line representations of molecules such as \gls{selfies} \cite{krenn2020self, krenn2022selfies} and \gls{smiles}.  
Taking this idea even further, \citet{Ramos2023} used this framework (with \gls{icl}) for Bayesian optimization---guiding experiments without even training models.

The projects in the following build on top of those initial results and extend them in  novel ways as well as by leveraging established techniques from quantum machine learning. 

Given that these encouraging results could be achieved with and without fine-tuning (i.e., updates to the weights of the model) for the language-interfaced training on tabular datasets, we use the term \gls{lift} also for \gls{icl} settings in which structured data is converted into text prompts for an \gls{llm}.

\paragraph{Molecular Energy Predictions}
A critical property in quantum chemistry is the atomization energy of a molecule, which gives us the basic thermochemical data used to determine a molecule's stability or reactivity. State-of-the-art quantum chemical methods (i.e., G4(MP2) \cite{Curtiss_2007}) can predict this energy with an accuracy of \SI{0.034}{\electronvolt} (or \SI{0.79}{\kilo cal \per \mol}) \cite{ramakrishnan2014quantum, narayanan2019accurate}. This accuracy is similar to, and in some cases even better than, the accuracy that can be reached experimentally. This motivated 
\citet{ramakrishnan2014quantum} and \citet{narayanan2019accurate} to compute these atomization energies for the 134,000 molecules in the QM9-G4MP2 dataset. 

The Berkeley-Madison team (Ankur~Gupta, Garrett~Merz, Alishba~Imran, and Wibe~de~Jong) used this dataset to fine-tune different \glspl{llm} using the \gls{lift} framework.
The team investigated if they could use an \gls{llm} to predict atomization energies with chemical accuracy. \citet{Jablonka_2023} emphasized that these \glspl{llm} might be particularly useful in the low-data limit. Here, we have a relatively large dataset, so it is an ideal system to gather insights into the performance of these models for datasets much larger than those used by \citet{Jablonka_2023}.

The Berkeley-Madison team showed that the \gls{lift} framework based on simple line representations such as \gls{smiles}  and \gls{selfies} \cite{krenn2020self, krenn2022selfies} can yield good predictions (R$^2>0.95$ on a holdout test set), that are, however, still inferior to dedicated models that have access to 3D information \cite{gupta2022three, ward2019machine}. 
An alternative approach to achieve chemical accuracy with \glspl{llm} tuned only on string representations is to leverage a $\Delta$-ML scheme \cite{Ramakrishnan_2015} in which the \gls{llm} is tuned to predict the difference between G4(MP2) and B3LYP \cite{Becke_1993} energies. \Cref{tab:qm9-lift} shows that good agreement could be achieved for the $\Delta$-ML approach. This showcases how techniques established for conventional \gls{ml} on molecules can also be applied with \glspl{llm}.
\begin{table*}
\footnotesize
    \caption{\emph{\Gls{lift} for molecular atomization energies on the QM9-G4MP2 dataset.} Metrics for models tuned on  90\% of the QM9-G4MP2 dataset (117,232 molecules), using 10\% (13,026 molecules) as a holdout test set. GPTChem refers to the approach reported by \citet{Jablonka_2023}, \gls{gpt}-2-\gls{lora} to \gls{peft} of the \gls{gpt}-2 model using \gls{lora}. The results indicate that the \gls{lift} framework can also be used to build predictive models for atomization energies, that can reach chemical accuracy using a $\Delta$-ML scheme. Baseline performance (mean absolute error reported by \citet{ward2019machine}): \SI{0.0223}{\electronvolt} for FCHL-based prediction of GP4(MP2) atomization energies and \SI{0.0045}{\electronvolt} (SchNet) and \SI{0.0052}{\electronvolt} (FCHL) for the $\Delta$-ML scheme.}
\begin{tabularx}{\textwidth}{p{12em}XXXX}
\toprule
mol. repr. \&  framework & \multicolumn{2}{l}{G4(MP2)

Atomization Energy} & \multicolumn{2}{l}{(G4(MP2)-B3LYP) Atomization Energy}\\
& R$^2$ & \gls{mad} / eV & R$^2$ & \gls{mad} / eV \\
\midrule
SMILES: GPTChem & 0.984 & 0.99 & 0.976 & 0.03 \\
SELFIES: GPTChem & 0.961 & 1.18 & 0.973 & 0.03 \\
SMILES: GPT2-LoRA & 0.931 & 2.03 & 0.910 & 0.06 \\
SELFIES: GPT2-LoRA & 0.959 & 1.93 & 0.915 & 0.06 \\ \bottomrule
\end{tabularx}
\label{tab:qm9-lift}
\end{table*}
 
Importantly, this approach is not limited to the OpenAI \gls{api}. With \glsfirst{peft} with \glsfirst{lora} \cite{hu2021lora} of the \gls{gpt}-2 model\cite{radford2019language}, one can also obtain comparable results on consumer hardware. These results make the \gls{lift} approach widely more accessible and allow research to the \gls{lift} framework for chemistry without relying on OpenAI. 

\paragraph{Text2Concrete} 
Concrete is the most used construction material, and the mechanical properties and climate impact of these materials are a complex function of the processing and formulation. Much research is focused on formulations of concrete that are less \ce{CO2} intensive.\cite{Scrivener_2018} 
To expedite the design process, e.g., by prioritizing experiments using \gls{ml}-predictions, data-driven methods have been investigated by \citet{Voelker2023} 
The Text2Concrete team (Sabine Kruschwitz, Christoph Völker, and Ghezal Ahmad Zia) explored, based on data reported by \citet{Rao2018}, whether \glspl{llm} can be used for this task. This data set provides 240 alternative, more sustainable, concrete formulations and their respective compressive strengths. 
From a practical point of view, one would like to have a model that can predict the compressive strength of the concrete as a function of its formulation. 

Interestingly, the largest \glspl{llm} can already give predictions without any fine-tuning. These models can \enquote{learn} from the few examples provided by the user in the prompt. 
Of course, such a few-shot approach (or \Gls{icl}, \cite{brown2020language}) does not allow for the same type of optimization as fine-tuning, and one can therefore expect it to be less accurate. However,  \citet{Ramos2023} showed that this method could perform well---especially if only so few data points are available such that fine-tuning is not a suitable approach. 

\begin{figure}
    \includegraphics[width=.8\textwidth]{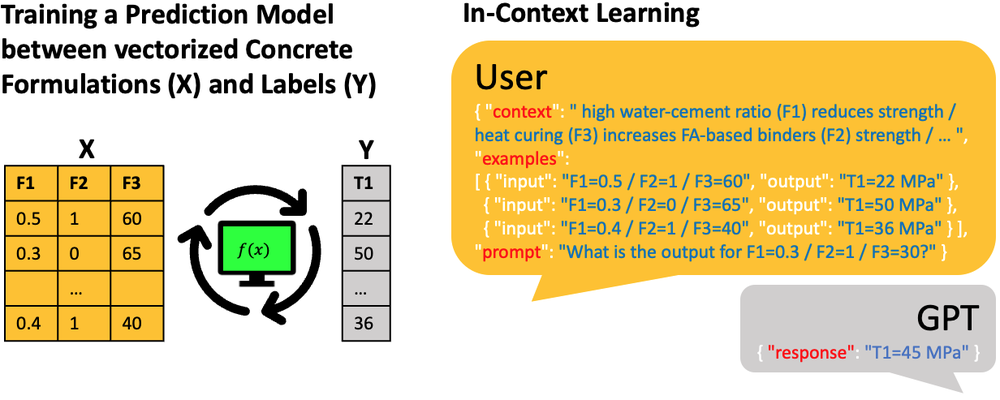}
    \caption{\emph{Using LLMs to predict the compressive strength of concretes}. An illustration of the conventional approach for solving this task, i.e., training classical prediction models using ten training data points as tabular data (left).  Using the \gls{lift} framework \glspl{llm} can also use tabular data and leverage context information provided in natural language (right). The context can be \enquote{fuzzy} design rules often known in chemistry and materials science but hard to incorporate in conventional \gls{ml} models. Augmented with this context and ten training examples, \gls{icl} with \gls{llm} leads to a performance that outperforms baselines such as \glspl{rf} or \gls{gpr}.}
    \label{fig:cement}
 \end{figure}

For their case study, the Text2Concrete team found a predictive accuracy comparable to a \gls{gpr} model (but inferior to a \gls{rf} model). 
However, one significant advantage of \glspl{llm} is that one can \emph{easily incorporate context}. The Text2Concrete team used this to include well-established design principles like the influence of the water-to-cement ratio on strength (\Cref{fig:cement}) into the modeling by simply stating the relationship between the features in natural language (e.g., \enquote{high water/cement ratio reduces strength}). 
This additional context reduced the outliers and outperformed the \gls{rf} model (R$^2$ of  0.67 and 0.72, respectively). 

The exciting aspect is that this is a typical example of domain knowledge that cannot be captured with a simple equation incorporable into conventional modeling workflows.
Such \enquote{fuzzy} domain knowledge, which may sometimes exist only in the minds of researchers, is common in chemistry and materials science.
With the incorporation of such \enquote{fuzzy} knowledge into \gls{lift}-based predictions using \glspl{llm}, we now have a novel and very promising approach to leverage such domain expertise that we could not leverage before. Interestingly, this also may provide a way to test \enquote{fuzzy} hypotheses, e.g., a researcher could describe the hypothesis in natural language and see how it affects the model accuracy. 
While the Text2Concrete example has not exhaustively analyzed how \enquote{fuzzy} context alterations affect \gls{llm} performance, we recognize this as a key area for future research that could enhance the application of \glspl{llm} and our approach to leveraging \enquote{fuzzy} domain knowledge within materials science.

\paragraph{Molecule Discovery by Context}
Much context is available in the full text of scientific articles.
This has been exploited by \citet{Tshitoyan_2019} who used a Word2Vec \cite{mikolov2013efficient} approach to embed words into a vector space. Word2Vec does so by tasking a model to predict for a word the probability for all possible next words in a vocabulary.
In this way, word embeddings capture syntactic and semantic details of lexical items (i.e., words).
When applied to material science abstracts, the word embeddings of compounds such as \ce{Li2CuSb} could be used for materials discovery by measuring their distance (cosine similarity) to concepts such as \enquote{thermoelectric}.\cite{Olivetti_2020}
However, traditional Word2Vec, as used by \citet{Tshitoyan_2019}, only produces \textit{static} embeddings, which remain unchanged after training.  
Word embeddings extracted from an \gls{llm}, on the other hand, are \textit{contextualized} on the specific sequence (sentence) in which they are used and, therefore, can more effectively capture the contexts of words within a given corpus \cite{selva2021review}.
Inspired by this, the GlobusLabs team (Zhi~Hong, Logan~Ward) investigated if similar embeddings could be used to discover hydrogen carrier molecules, that are relevant for energy storage applications. 
For this, they leverage the ScholarBert model \cite{hong2022scholarbert} trained on a large corpus of scientific articles collected by the Public.Resource.Org nonprofit organization. For different candidate molecules, they searched for sentences in the Public.Resource.Org corpus and used the average of the embeddings of these sentences as a fingerprint of the molecules.
Given those fingerprints, they could rank molecules by how close their fingerprints are to the ones of known hydrogen carrier molecules. Visual inspection indicates that the selected molecules indeed bear similarities to known hydrogen carrier molecules.

\paragraph{Text template paraphrasing}
In the \gls{lift} framework used in the examples above, the data are embedded in so-called prompt templates that can have a form like \texttt{What is the <property name> of <representation>?}, where the texts in chevrons are placeholders that are replaced with actual values such as \enquote{solubility} and \enquote{2-acetyloxybenzoic acid}.
In the low-data regime, data points are \enquote{wasted} by the model needing to learn the syntax of the prompt templates. In the big-data regime, in contrast, one might worry that the model loses some of its general language modeling abilities by always dealing with the same template.
This naturally raises the question if one can augment the dataset to mitigate these problems---thereby leveraging again, similar to $\Delta$-\gls{ml}, a technique that has found use in conventional \gls{ml} previously.
However, text-based data are challenging to augment due to their discrete nature and the fact that the augmented text still needs to be syntactically and semantically valid. 
Interestingly, as Michael Pieler (OpenBioML.org and Stability.AI) shows (and as has been explored by \citet{dai2023auggpt}), it turns out that \glspl{llm} can also be used to address this problem by simply prompting an \gls{llm} (e.g., \gls{gpt}-4 or Anthrophic's Claude) to paraphrase a prompt template (see SI section ID). 

This approach will allow us to automatically create new paraphrased high-quality prompts for \gls{lift}-based training very efficiently---to augment the dataset and reduce the risk of overfitting to a specific template. Latter might be particularly  important if one still wants to retain general language abilities of the \glspl{llm} after finetuning.

\paragraph{Genetic algorithm using an \gls{llm}}
Genetic algorithms are popular methods for generating new structures; they are evolutionary algorithms in which building blocks (e.g., fragments of \gls{smiles} strings) are iteratively crossed over, mutated, and subjected to other genetic operations to evolve structures with better performance (such as catalysts with higher conversion) \cite{venkatasubramanian1994computer}. 
The efficiency of such a genetic algorithm often depends on how well the genes and genetic operations match the underlying chemistry. For example, if the algorithm replaces atom by atom, it may take several generations before a complete functional group is replaced.  

One might hypothesize that \glspl{llm} can make the evolution process more efficient, e.g., by using an \gls{llm} to handle the reproduction. One might expect that inductive biases in the \gls{llm} help create recombined molecules which are more chemically viable, maintaining the motifs of the two parent molecules better than a random operation.

 \begin{figure}
    \includegraphics[width=\textwidth]{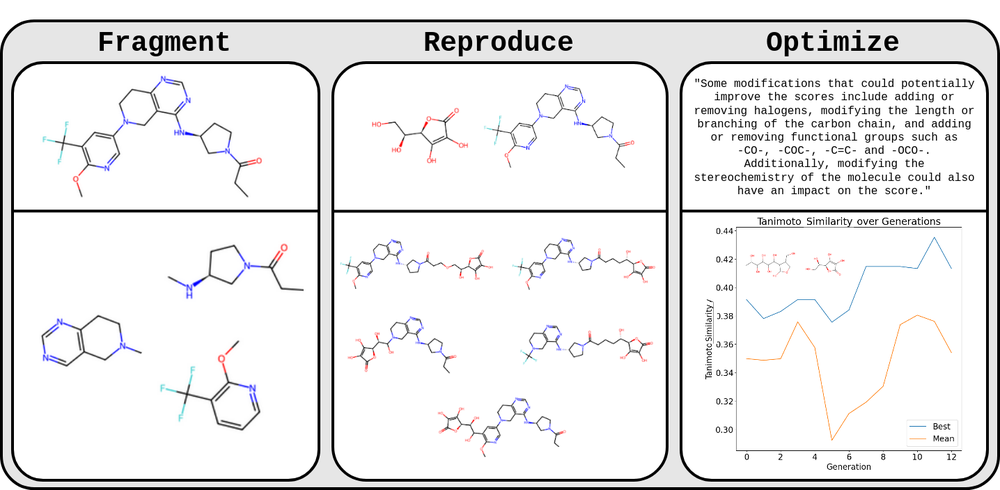}
    \caption{\emph{\gls{ga} using an \gls{llm}}. This figure illustrates how different aspects of a \gls{ga} can be performed by an \gls{llm}. \gls{gpt}-3.5 was used to fragment, reproduce, and optimize molecules represented by \gls{smiles} strings.
    The first column illustrated how  an \gls{llm} can fragment a molecule represented by a \gls{smiles} string (input molecule on top, output \gls{llm} fragments below). 
    The middle column showcases how an \gls{llm} can reproduce/mix two molecules as is done in a \gls{ga} (input molecule on top, output \gls{llm} below). 
    The right column illustrates an application in which an \gls{llm} is used to optimize molecules given their \gls{smiles} and an associated score. The \gls{llm} suggested potential modifications to optimize molecules. The plot shows best (blue) and mean (orange) Tanimoto similarity to Vitamin C per \gls{llm} produced generations.}
    \label{fig:llmga}
 \end{figure} 
 
The team from McGill University (Benjamin~Weiser, Jerome~Genzling, Nicolas~Gastellu, Sylvester~Zhang, Tao~Liu, Alexander~Al-Feghali, Nicolas~Moitessier) set out the first steps to test this hypothesis (\Cref{fig:llmga}). In initial experiments, they found that \gls{gpt}-3.5, without any finetuning, can fragment molecules provided as \gls{smiles} at rotatable bonds with a success rate of \SI{70}{\percent}. 
This indicates that \gls{gpt}-3.5 understands \gls{smiles} strings and aspects of their relation to the chemical structures they represent. 
Subsequently, they asked the \glspl{llm} to fragment and recombine two given molecules. The \gls{llm} frequently created new combined molecules with fragments of each species which were reasonable chemical structures more often than a random \gls{smiles} string combining operation (two independent organic chemists judged the \gls{llm}-\gls{ga}-generated molecules to be chemically reasonable in \nicefrac{32}{32} cases, but only in \nicefrac{21}{32} cases for the random recombination operation).

Encouraged by these findings, they prompted an \gls{llm} with 30 parent molecules and their performance scores (Tanimoto similarity to vitamin C) with the task to come up with $n$ new molecules that the \gls{llm} \enquote{believes} to improve the score. A preliminary visual inspection suggests that the \gls{llm} might produce chemically reasonable modifications. Future work will need to systematically investigate potential improvements compared to conventional \glspl{ga}.

The importance of the results of the McGill team is that they indicate that these \glspl{llm} (when suitably conditioned) might not only reproduce known structures but generate new structures that make chemical sense\cite{flam-shepherd2023language, Jablonka_2023}. 

A current limitation of this approach is that most \glspl{llm} still struggle to output valid \gls{smiles} without explicit fine-tuning \cite{white2023assessment}. 
We anticipate that this problem might be mitigated by building foundation models for chemistry (with more suitable tokenization \cite{taylor2022galactica, Schwaller_2018}), as, for instance, the ChemNLP project of OpenBioML.org attempts to do (\url{https://github.com/OpenBioML/chemnlp}). 
In addition, the context length limits the number of parent molecules that can be provided as examples.

Overall, we see that the flexibility of the natural language input and the in-context learning abilities allows using \glspl{llm} in very different ways---to very efficiently build predictive models or to approach molecular and material design in entirely unprecedented ways, like by providing context---such as \enquote{fuzzy} design rules---or simply prompting the \gls{llm} to come up with new structures.
However, we also find that some \enquote{old} ideas, such as $\Delta$-ML and data augmentation, can also be applied in this new paradigm.

\subsection{Automation and novel interfaces}
\citet{yao2023react} and \citet{schick2023toolformer} have shown that \glspl{llm} can be used as agents that can autonomously make use of external tools such as Web-\glspl{api}---a paradigm that some call MRKL (pronounced \enquote{miracle}) Systems---modular reasoning, knowledge, and language systems \cite{karpas2022mrkl}. 
By giving \glspl{llm} access to tools and forcing them to think step-by-step \cite{wei2022chainofthought}, we can thereby convert \glspl{llm} from hyperconfident models that often hallucinate to systems that can reason based on observations made by querying robust tools. As the technical report for \gls{gpt}-4 highlighted\cite{openai2023gpt4}, giving \glspl{llm} access to tools can lead to emergent behavior, i.e., enabling the system to do things that none of its parts could do before.
In addition, this approach can make external tools more accessible---since users no longer have to learn tool-specific \glspl{api}. It can also make tools more interoperable---by using natural language instead of \enquote{glue code} to connect tools. 

This paradigm has recently been used by \citet{bran2023chemcrow} to create digital assistants that can call and combine various tools such as Google search and the IBM RXN retrosynthesis tool when prompted with natural language. \citet{boiko2023emergent} used a similar approach and gave \glspl{llm}  access to laboratories via  cloud lab \glspl{api}. In their system, the \gls{llm} could use external tools to plan a synthesis, which it could execute using the cloud lab.  

\paragraph{MAPI-LLM}
Electronic structure calculations have reached such a high level of accuracy that one can answer questions like \enquote{Is the material AnByCz stable?} 
Indeed, the Materials Project\cite{Jain2013} stores thermodynamic data on many components from which one can obtain a reasonable estimate of the stability of a given material. 
Or, if the material is not in the database, one can do a simulation instead. 
Similarly, to answer prompts such as \enquote{Give me a reaction to produce CaCO3}, there is a lot of helpful information in the Materials Project database and the internet that can help to come up with an answer. 

To answer these questions, state-of-the-art computational tools or existing databases can be used. However, their use often requires expert knowledge. To use existing databases, one must choose which database to use, how to query the database, and what representation of the compound is used (e.g., \gls{inchi}, \gls{smiles}, etc.). Otherwise, if the data is not in a database, one must run calculations, which requires a deep understanding of technical details. 
\Glspl{llm} can simplify the use of such tools. By typing in a question, we can prompt the \gls{llm} to translate this question into a workflow that leads to the answer.

The MAPI-LLM team (Mayk Caldas Ramos, Sam Cox, Andrew White) made the first steps towards developing such a system (MAPI-LLM) and created a procedure to convert a text prompt into a query of the \gls{mapi} to answer questions such as \enquote{Is the material AnByCz stable?}
In addition, MAPI-LLM is capable of handling classification queries, such as \enquote{Is Fe2O3 magnetic?}, as well as regression problems, such as \enquote{What is the band gap of Mg(Fe2O3)2?}.

Because an \gls{llm} is used to create the workflow, MAPI-LLM can process even more complex questions.
For instance, the question \enquote{If Mn23FeO32 is not metallic, what is its band gap?} should create a two-step workflow first to check if the material is metallic and then calculate its band gap if it is not.

Moreover, MAPI-LLM applies \gls{icl} if the data for a material's property is unavailable via the \gls{mapi}. MAPI-LLM generates an \gls{icl} prompt,  building context based on the data for similar materials available in Materials Project database. This context is then leveraged by an \gls{llm} to infer properties for the unknown material. This innovative use of \gls{icl} bridges data gaps and enhances MAPI-LLM's robustness and versatility.

\begin{figure}
    \centering
    \includegraphics[width=.8\textwidth]{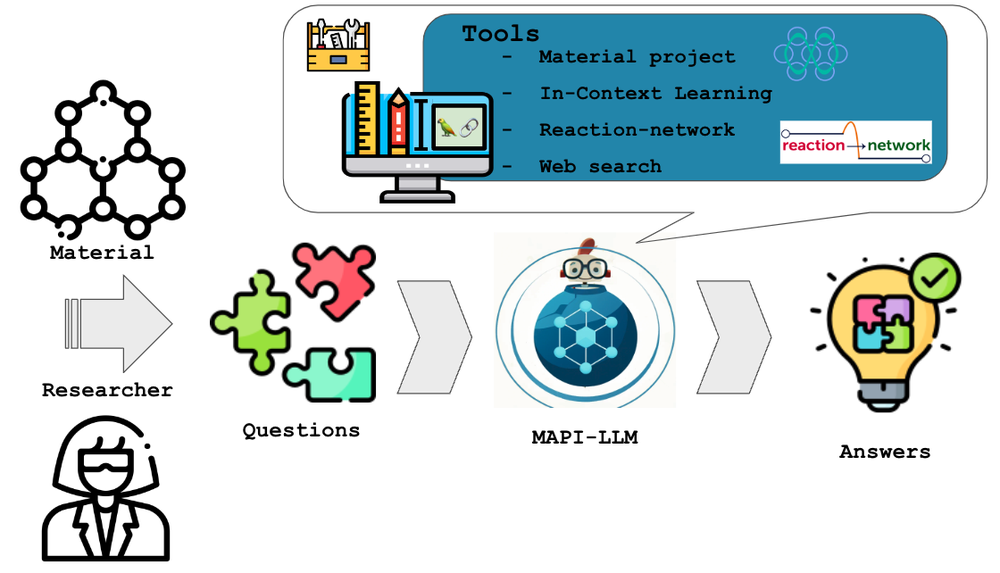}
    \caption{\emph{Schematic overview of the MAPI-LLM workflow.} It uses \glspl{llm} to process the user's input and decide which available tools (e.g., Materials Project \gls{api}, the Reaction-Network package, and Google Search) to use following an iterative chain-of-thought procedure. In this way, it can answer questions such as \enquote{Is the material AnByCz stable?}.}
    \label{fig:mapi}
\end{figure}

\paragraph{sMolTalk}
The previous application already touches on the problem that software for chemical applications requires scientists to invest a significant amount of time in learning even the most basic applications. An example of this is visualization software.   
 Depending on the package and its associated documentation, chemists and materials scientists might spend hours to days learning the details of specific visualization software that is sometimes poorly documented. And in particular, for occasional use, if it takes a long time to learn the basics, it won't be used.

\begin{figure} 
    \centering
    \includegraphics[width=.7\textwidth]{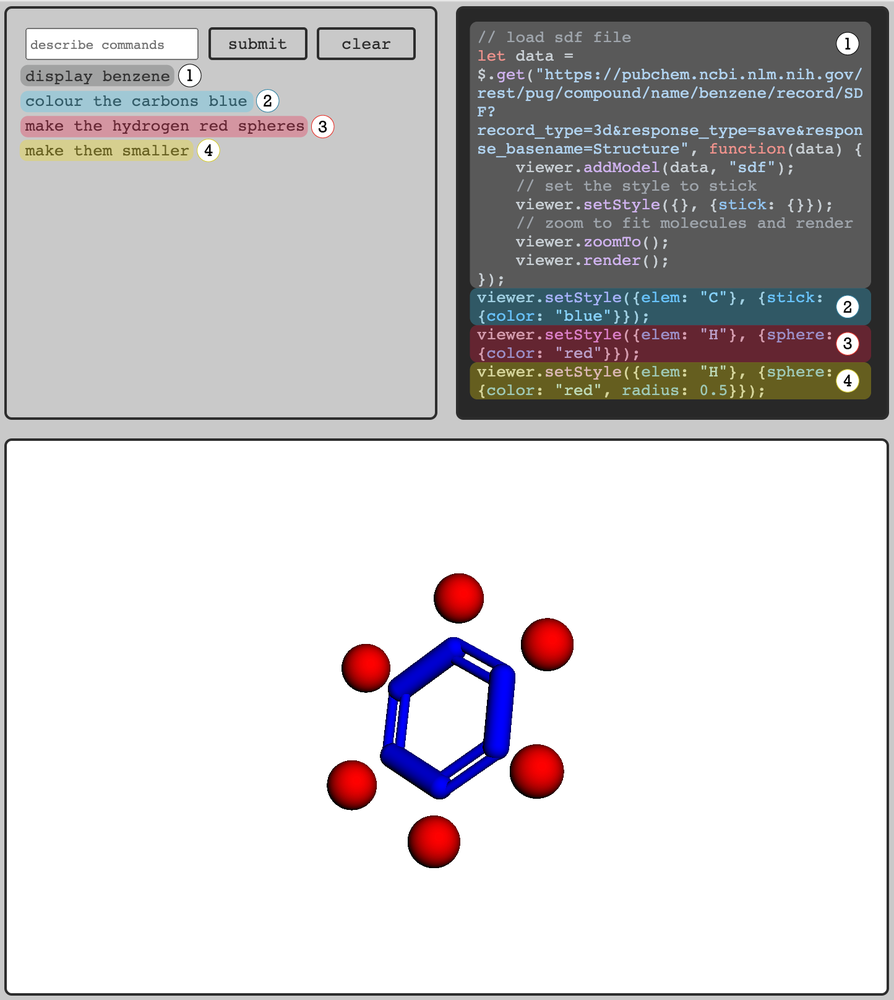}
    \caption{\emph{The sMolTalk interface.} Based on few-shot prompting \glspl{llm} can create code for visualization tools such as \texttt{3dmol.js} that can create custom visualization based on a natural-language description of the desired output.
    The top left box is the input field where users can enter commands in natural language. The top right box prints the code the \gls{llm} generates. This code generates the visualization shown in the lower box. In this example, the user entered a sequence of four commands: the \gls{llm} (1) generates code for retrieving the structure, (2) colors the carbons blue, (3) displays the hydrogens as red spheres,  and (4) reduces the size of the spheres.
    } 
    \label{fig:sMolTalk}
\end{figure}

As the sMolTalk-team (Jakub~Lála, Sean~Warren, Samuel~G.~Rodriques) showed, one can use \glspl{llm} to write code for visualization tools such as \texttt{3dmol.js} to address this inefficiency~\cite{Rego_2014}. 
Interestingly, few-shot prompting with several examples of user input with the expected JavaScript code that manipulates the \texttt{3dmol.js} viewer is all that is needed to create a prototype of an interface that can retrieve protein structures from the \gls{pdb} and create custom visualization solutions, e.g., to color parts of a structure in a certain way (\Cref{fig:sMolTalk}). 
The beauty of the language models is that the user can write the prompt in many different (\enquote{fuzzy}) ways: whether one writes \enquote{color} or \enquote{colour}, or terms like \enquote{light yellow} or \enquote{pale yellow} the \gls{llm} translates it into something the visualization software can interpret.

However, this application also highlights that further developments of these \gls{llm}-based tools are needed. For example, a challenge the sMolTalk tool faces is robustness. 
For instance, fragments from the prompt tend to leak into the output and must be handled with more involved mechanisms, such as retries in which one gives the \glspl{llm} access to the error messages or prompt engineering.  
Further improvement can also be expected if the application leverages a knowledge base such as the documentation of \texttt{3dmol.js}. 

As the work of Glenn Hocky and Andrew White shows \cite{marvis}, an \gls{llm}-interface for software can also be used with other programs such as \texttt{VMD} \cite{Humphrey_1996} and extended with speech-to-text models (such as Whisper \cite{radford2022robust}) to enable voice control of such programs.
In particular, such an \gls{llm}-based agent approach might be implemented for the \texttt{PyMOL} program, where various tools for protein engineering could be interfaced through a chat interface, lowering the barrier to entry for biologists to use recent advancements within \textit{in silico} protein engineering (such as RosettaFold \cite{Baek_2021} or RFDiffusion \cite{Watson_2022}).

\begin{figure}
    \centering
    \includegraphics[width=.6\textwidth]{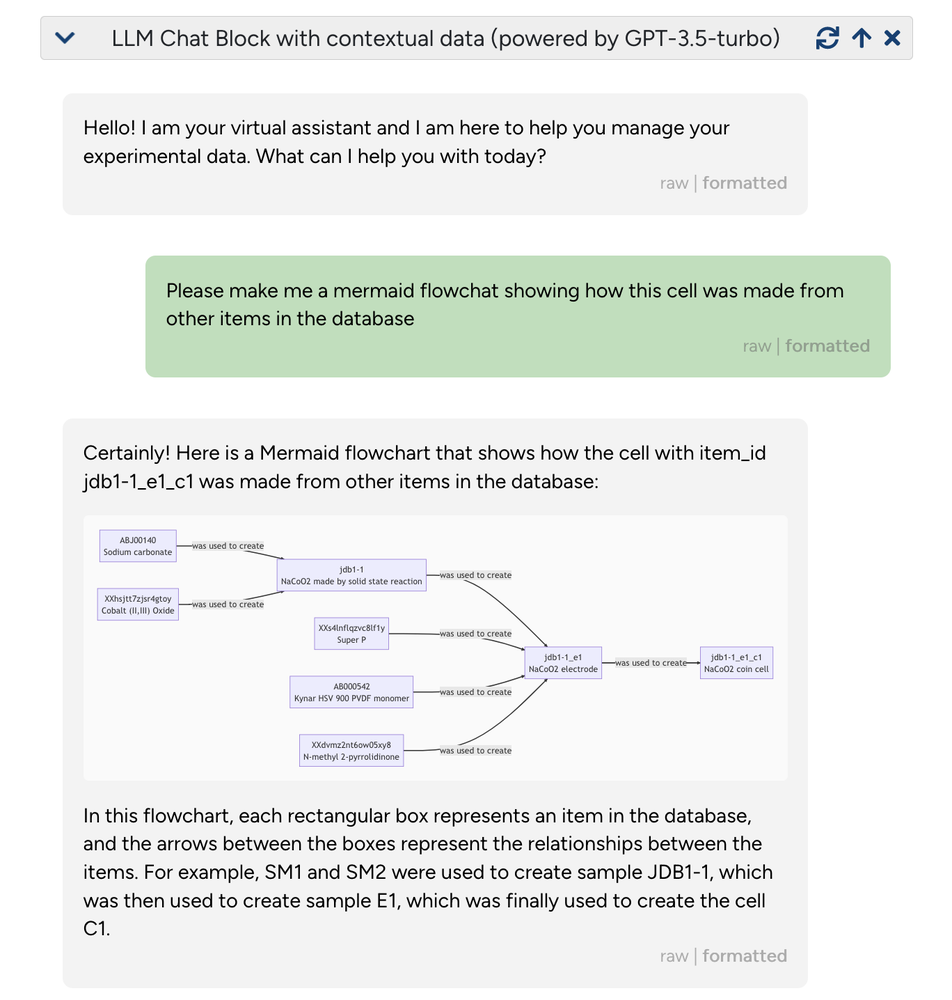}
    \caption{\emph{Using an LLM as an interface to an \gls{eln}/data management system.} \gls{llm}-based assistants can provide powerful interfaces to digital experimental data. The figure shows a screenshot of a conversation with \texttt{whinchat} in the \texttt{datalab} data management system (\url{https://github.com/the-grey-group/datalab}). Here, \texttt{whinchat} is provided with data from the \gls{json} \gls{api} of \texttt{datalab} of an experimental battery cell. The user then prompts (green box) the system to build a flowchart of the provenance of the sample. The assistant responds with \texttt{mermaid.js} markdown code, which the \texttt{datalab} interface automatically recognizes and translates into a visualization. }
    \label{fig:datalab_assistant}
\end{figure}

\paragraph{\gls{eln} interface: \texttt{whinchat}}
In addition to large, highly curated databases with well-defined data models \cite{Andersen2021} (such as those addressed by the MAPI-LLM project), experimental materials and chemistry data is increasingly being captured using digital tools such as \glspl{eln} and or \glspl{lims}.
Importantly, these tools can be used to record both structured and unstructured lab data in a manner that is actionable by both humans and computers.
However, one challenge in developing these systems is that it is difficult for a traditional user interface to have enough flexibility to capture the richness and diversity of real, interconnected, experimental data. 
Interestingly, \glspl{llm} can interpret and contextualize both structured and unstructured data and can therefore be used to create a novel type of flexible, conversational interface to such experimental data. The \texttt{whinchat} team (Joshua~D.~Bocarsly, Matthew~L.~Evans, and Ben~E.~Smith) embedded an \gls{llm} chat interface within \texttt{datalab}, an open source materials chemistry data management system, where the virtual \gls{llm}-powered assistant can be \enquote{attached} to a given sample. 
The virtual assistant has access to responses from the \gls{json} \gls{api} of \texttt{datalab} (containing both structured and unstructured/free text data) and can use them to perform several powerful tasks: 
First, it can contextualize existing data by explaining related experiments from linked responses, resolving acronyms/short-hand notations used by experimentalists, or creating concise textual summaries of complex and nested entries. 
Second, it can reformat or render the data, for instance, by creating (\texttt{mermaid.js}) flowcharts or (Markdown) tables (\Cref{fig:datalab_assistant}). 
Third, it can use its generic reasoning abilities to suggest future experiments, for instance, related materials to study, synthesis protocols to try, or additional characterization techniques. 
This is shown in the examples given in SI section 2C, where \texttt{whinchat} was able to provide hints about which NMR-active nuclei can be probed in the given sample.

It is easy to envision that this tool could be even more helpful by fine-tuning or conditioning it on a research group's knowledge base (e.g., group Wiki or standard operating procedures) and communication history (e.g., a group's Slack history). 
An important limitation of the current implementation is that the small context window of available \glspl{llm} limits the amount of \gls{json} data one can directly provide within the prompt, limiting each conversation to analyzing a relatively small number of samples. 
Therefore, one needs to either investigate the use of embeddings to determine which samples to include in the context or adopt an \enquote{agent} approach where the assistant is allowed to query the \gls{api} of the \gls{eln} (interleaved with extraction and summarization calls).

\begin{figure}
    \centering
    \includegraphics[width=.6\textwidth]{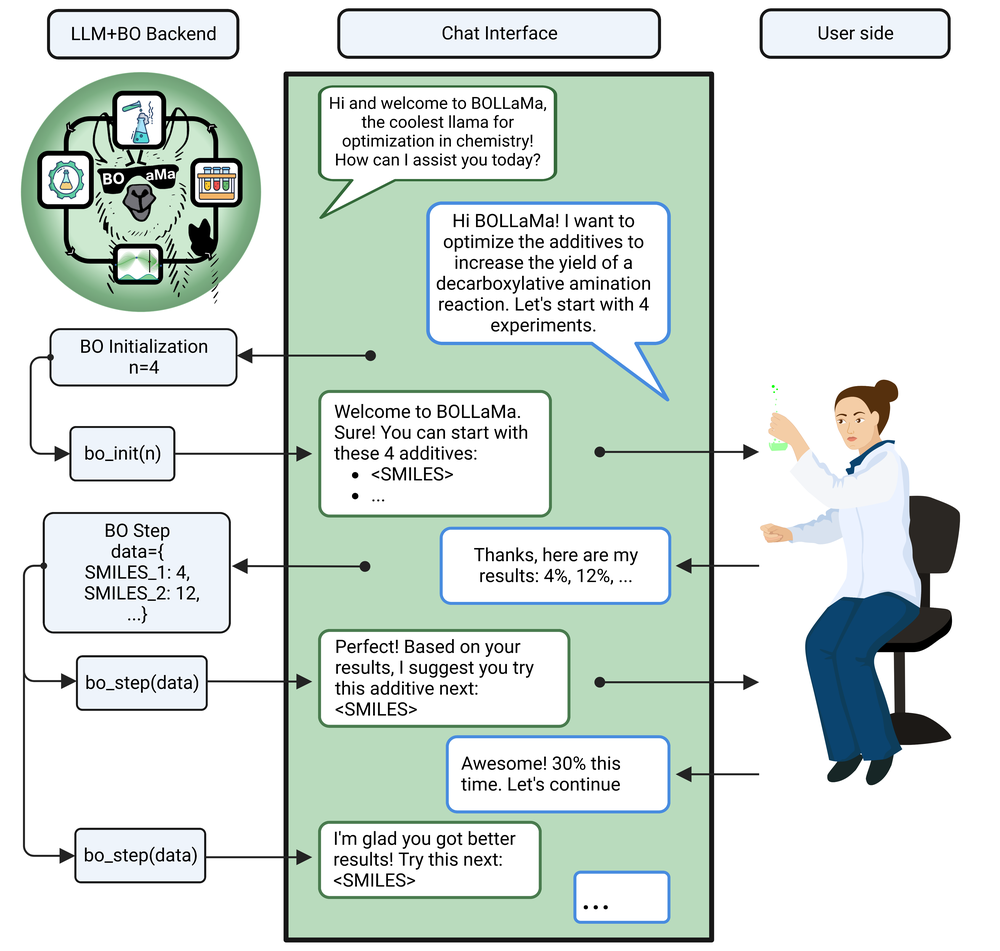}
    \caption{\emph{Schematic overview of BoLLama.} An \gls{llm} can act as an interface to a \gls{bo} algorithm. An experimental chemist can bootstrap an optimization and then, via a chat interface, update the state of the simulation to which the bot responds with the recommended next steps.  }
    \label{fig:bollama}
\end{figure}

\paragraph{BOLLaMa: facilitating Bayesian optimization with large language models}
\Gls{bo} is a powerful tool for optimizing expensive functions, such as mapping of reaction conditions to the reaction yield. 
Chemists would greatly benefit from using this method to reduce the number of costly experiments they need to run\cite{volk_alphaflow_2023, shields_bayesian_2021}.
However, \gls{bo} faces an interface and accessibility problem, too. 
The existing frameworks require significant background knowledge and coding experience not conventionally taught in chemistry curricula. Therefore, many chemists cannot benefit from tools such as \gls{bo}.   
The BOLLaMa-team (Bojana Ranković, Andres M. Bran, Philippe Schwaller) showed that \Glspl{llm} can lower the barrier for the use of \gls{bo} by providing a natural language chat-like interface to \gls{bo} algorithms. 
\Cref{fig:bollama} shows a prototype of a chat interface in which the \gls{llm} interprets the user request, initializes a \gls{bo} run by suggesting initial experimental conditions, and then uses the feedback of the user to drive the \gls{bo} algorithm and suggest new experiments. The example used data on various additives for a cooperative nickel-photoredox catalyzed reaction \cite{prieto2022accelerating} and the \gls{bo} code from \citet{rankovic_bayesian_2022}. 
This ideally synergizes with an \gls{llm} interface to a data management solution (as discussed in the previous project) as one could directly persist the experimental results and leverage prior records to \enquote{bootstrap} \gls{bo} runs. 

As the examples in this section show, we find that \glspl{llm} have the potential to greatly enhance the efficiency of a diverse array of processes in chemistry and materials science by providing novel interfaces to tools or by completely automating their use. This can help streamline workflows, reduce human error, and increase productivity---often by replacing \enquote{glue code} with natural language or studying a software library by chatting with an \gls{llm}.

\subsection{Knowledge Extraction}
Beyond proving novel interfaces for tools, \glspl{llm} can also serve as powerful tools for extracting knowledge from the vast amount of chemical literature available. 
With \glspl{llm}, researchers can rapidly mine and analyze large volumes of data, enabling them to uncover novel insights and advance the frontiers of chemical knowledge. 
Tools such as paper-qa \cite{paperqa} can help to dramatically cut down the time required for literature search by automatically retrieving, summarizing, and contextualizing relevant fragments from the entire corpus of the scientific literature---for example, answering questions (with suitable citations) based on a library of hundreds of documents \cite{White_2023}.
As the examples in the previous section indicated, this is particularly useful if the model is given access to search engines on the internet. 

\paragraph{InsightGraph}
To facilitate downstream use of the information,  \Glspl{llm} can also convert unstructured data---the typical form of these literature reports---into structured data.
The use of \gls{gpt} for this application has been reported by \citet{dunn2022structured} and \citet{walker2023extracting}, who used an iterative fine-tuning approach to extract data structured in \gls{json} from papers. 
In their approach, initial (zero-shot) completions of the \gls{llm} are corrected by domain experts. Those corrected completions are then used to finetune \glspl{llm}, showing improved performance on this task.

\begin{figure*}
    \centering
    \includegraphics[width=.8\textwidth]{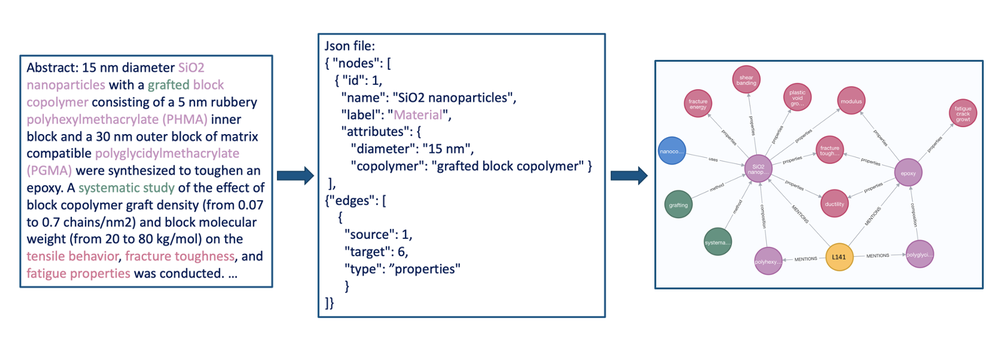}
    \caption{\emph{The InsightGraph interface.} A suitably prompted \gls{llm} can create knowledge graph representations of scientific text that can be visualized using tools such as neo4j's visualization tools. \cite{neo4jpackage} }
    \label{fig:InsightGraph}
\end{figure*}

However, for certain applications, one can construct powerful prototypes using only careful prompting. For instance, the InsightGraph team (Defne~Circi, Shruti~Badhwar) showed that \gls{gpt}-3.5-turbo, when prompted with an example \gls{json} containing a high-level schema and information on possible entities (e.g., materials) and pairwise relationships (e.g., properties), can, as \Cref{fig:InsightGraph} illustrates, provide a knowledge graph representation of the entities and their relationships in a text describing the properties and composition of polymer nanocomposites. 
A further optimized version of this tool might offer a concise and visual means to quickly understand and compare material types and uses across sets of articles and could be used to launch a literature review. An advanced potential application is the creation of structured, materials-specific datasets for fact-based question-answering and downstream machine-learning tasks.

\begin{figure}
    \centering
    \includegraphics[width=.48\textwidth]{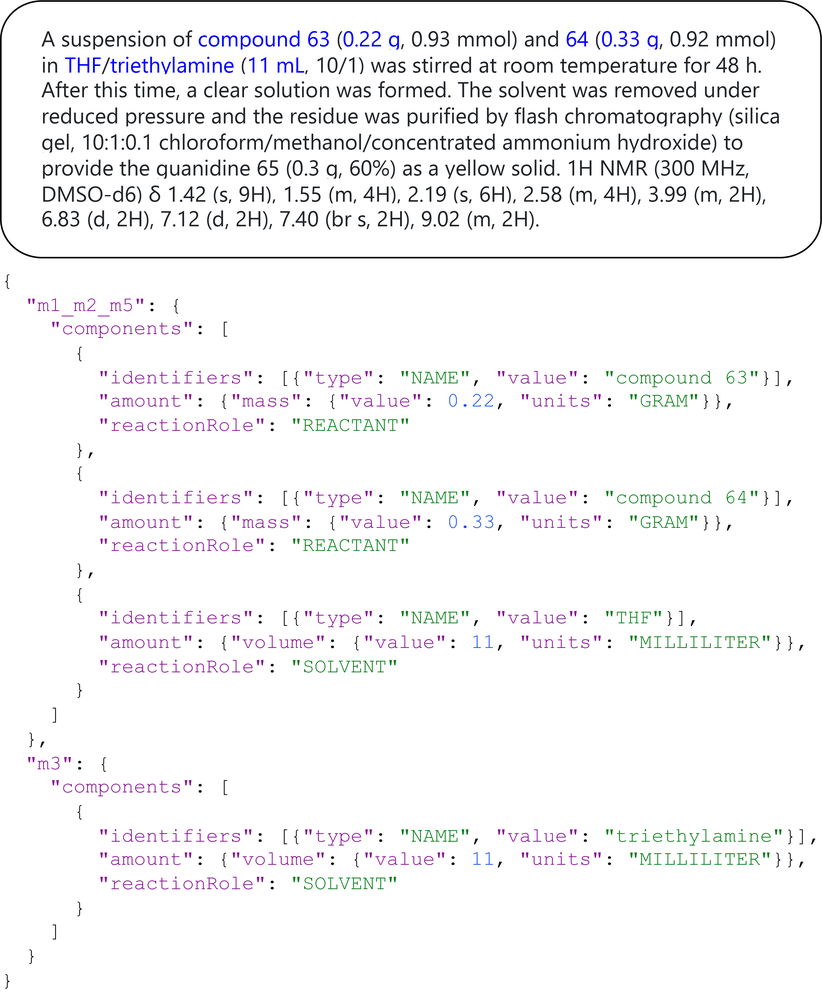}
    \caption{\emph{The Organic Synthesis Parser interface.} The top box shows text describing an organic reaction (\url{https://open-reaction-database.org/client/id/ord-1f99b308e17340cb8e0e3080c270fd08}), which the finetuned \gls{llm} converts into structured \gls{json} (bottom). A demo application can be found at \url{https://qai222.github.io/LLM_organic_synthesis/}.
}
    \label{fig:organic-synthesis-parser}
\end{figure}

\paragraph{Extracting Structured Data from Free-form Organic Synthesis Text}
Unstructured text is commonly used for describing organic synthesis procedures. 
Due to the large corpus of literature, manual conversion from unstructured text to structured data is unrealistic. However, structured data are needed for building conventional \gls{ml} models for reaction prediction and condition recommendation.
The \gls{ord}\cite{kearnes_open_2021} is a database of curated organic reactions. In the \gls{ord}, while reaction data are structured by the \gls{ord} schema, many of their procedures are also available as plain text. Interestingly, an \gls{llm} (e.g., OpenAI's \texttt{text-davinci-003}) can, after finetuning on only 300 prompt-completion pairs, extract \SI{93}{\percent} of the components from the free-text reaction description into valid \glspl{json} (\Cref{fig:organic-synthesis-parser}). Such models might significantly increase the data available for training models on tasks such as predicting reaction conditions and yields. 
It is worth noting that all reaction data submitted to \gls{ord} are made available under the CC-BY-SA license, which makes \gls{ord} a suitable data source for fine-tuning or training an \gls{llm} to extract structured data from organic procedures. 
A recent study on gold nanorod growth procedures also demonstrated the ability of \gls{llm} in a similar task.\cite{walker2023extracting}
In contrast to the \gls{lift}-based prediction of atomization energies reported in the first section by the Berkeley-Madison team, parameter-efficient fine-tuning of the open-source Alpaca model\cite{stanford_alpaca, alpaca_lora, touvron2023llama} using \gls{lora}\cite{hu2021lora} did not yield a model that can construct valid \glspl{json}.

\paragraph{TableToJson: Structured information from tables in scientific papers}
The previous example shows how structured data can be extracted from plain text using \glspl{llm}. However, relevant information in the scientific literature is not only found in text form. 
Research papers often contain tables that collect data on material properties, synthesis conditions, and results of characterization and experiments.
Converting table information into structured formats is essential to enable automated data analysis, extraction, and integration into computational workflows.
Although some techniques could help in the process of extracting this information (performing OCR or parsing XML), converting this information in structured data following, for example, a specific \gls{json} schema with models remains a challenge. 
The INCAR-CSIC team showed that the OpenAI \texttt{text-davinci-003} model, when prompted with a desired \gls{json} schema and the \gls{html} of a table contained in a scientific paper, can generate structured \gls{json} with the data in the table.

First, the OpenAI \texttt{text-davinci-003} model was directly used to generate \gls{json} objects from the table information. 
This approach was applied to several examples using tables collected from papers on different research topics within the field of chemistry\cite{GhanbarpourMamaghani2023,Peng2013,Sahoo2023,Suppiah2021,GonzlezVzquez2018,Mohsin2023,Kaur2020}.
The accuracy for those different examples, calculated as the percentage of schema values generated correctly, is shown in \Cref{fig:tablesjson_main}.
When the OpenAI model was prompted with the table and desired schema to generate a \gls{json} object, it worked remarkably well in extracting the information from each table cell and inserting it at the expected place in the schema. 
As output, it provided a valid \gls{json} object with a \SI{100}{\percent} success rate of error-free generated values in all the studied examples. 
However, in some examples, the model did not follow the schema.

To potentially address this problem the team utilized the  \texttt{jsonformer} approach. This tool reads the keys from the \gls{json} schema and only generates the value tokens, guaranteeing the generation of a syntactically valid \gls{json} (corresponding to the desired schema) by the \gls{llm}\cite{jsonformer,jsonformeropenai}. Using an \gls{llm} without such a decoding strategy cannot guarantee that valid \gls{json} outputs are produced. 
With the \texttt{jsonformer} approach, in most cases, by using a simple descriptive prompt about the type of input text, structured data can be obtained with \SI{100}{\percent} correctness of the generated values. 
In one example, an accuracy of \SI{80}{\percent} was obtained due to errors in the generation of numbers in scientific notation.
For a table with more complex content (long molecule names, hyphens, power numbers, subscripts, and superscripts,\dots) the team achieved an accuracy of only \SI{46}{\percent}.
Most of these issues could be solved by adding a specific explanation in the prompt, increasing the accuracy to \SI{100}{\percent} in most cases.

Overall, both approaches performed well in generating the \gls{json} format. 
The OpenAI \texttt{text-davinci-003} model could correctly extract structured information from tables and give a valid \gls{json} output, but it cannot guarantee that the outputs will always follow the provided schema.
\texttt{Jsonformer} may present problems when special characters need to be generated, but most of these issues could be solved with careful prompting. 
These results show that \glspl{llm} can be a useful tool to help to extract scientific information in tables and convert it into a structured form with a fixed schema that can be stored in a database, which could encourage the creation of more topic-specific databases of research results.

\begin{figure}[htb]
    \centering
    \includegraphics[width=\textwidth]{figures/tablesjson/tablesjson_main_text.pdf}
    \caption{\emph{TableToJson.} Results of the structured \gls{json} generation of tables contained in scientific articles. Two approaches are compared: (i) the use of an OpenAI model prompted with the desired \gls{json} schema, and (ii) the use of an OpenAI model together with \texttt{jsonformer}. In both cases, \gls{json} objects were always obtained. The output of the OpenAI model did not always follow the provided schema, although this might be solved by modifying the schema. 
    The accuracy of the results from the \texttt{jsonformer} approach used with OpenAI models could be increased (as shown by the blue arrows) by solving errors in the generation of power numbers and special characters with a more detailed prompt.
    The results can be visualized in this demo app: \url{https://vgvinter-tabletojson-app-kt5aiv.streamlit.app/}}    
    \label{fig:tablesjson_main}
\end{figure}

\paragraph{AbstractToTitle \& TitleToAbstract: text summarization and text generation}
Technical writing is a challenging task that often requires presenting complex abstract ideas in limited space. For this, frequent rewrites of sections are needed, in which \glspl{llm} could assist domain experts. 
Still, evaluating their ability to generate text such as a scientific paper is essential, especially for chemistry and materials science applications.

Large datasets of chemistry-related text are available from open-access platforms such as arXiv and PubChem. These articles contain titles, abstracts, and often complete manuscripts, which can be a testbed for evaluating \glspl{llm} as these titles and abstracts are usually written by expert researchers. 
Ideally, an \gls{llm} should be able to generate a title of an abstract close to the one developed by the expert, which can be considered a specialized text-summarization task. 
Similarly, given a title, an  \gls{llm} should generate text close to the original abstract of the article, which can be considered a specialized text-generation task. 


These tasks have been introduced by the AbstractToTitle \& TitleToAbstract team (Kamal Choudhary) in the JARVIS-ChemNLP package\cite{choudhary2022chemnlp}. 
For text summarization, it uses a pre-trained Text-to-Text Transfer Transformer (T5) model developed by Google\cite{raffel2020exploring} that is further fine-tuned to produce summaries of abstracts. 
On the arXiv condensed-matter physics (cond-mat) data, the team found that fine-tuning the model can help improve the performance (\gls{rouge}-1 score of \SI{39.0}{\percent} which is better than an untrained model score of \SI{30.8}{\percent} for an 80/20 split).

For text generation, JARVIS-ChemNLP finetunes the pretrained \gls{gpt}-2-medium \cite{radford2019language} model available in the HuggingFace library.\cite{Wolf_2020} After finetuning, the team found a \gls{rouge} score of \SI{31.7}{\percent}, which is a good starting point for pre-suggestion text applications. 
Both tasks with well-defined train and test splits are now available in the JARVIS-Leaderboard platform for the AI community to compare other \glspl{llm} and systematically improve the performance.

In the future, such title to abstract capabilities can be extended to generating full-length drafts with appropriate tables, multi-modal figures, and results as an initial start for the human researcher to help in the technical writing processes. Note that there have been recent developments in providing guidelines for using \gls{llm}-generated text in technical manuscripts\cite{editorials2023tools}, so such an \gls{llm} model should be considered as an assistant of writing and not the master/author of the manuscripts.


\subsection{Education}

Given all the opportunities \gls{llm} open for materials science and chemistry, there is an urgent need for education to adapt. 
Interestingly, \glspl{llm} also provide us with entirely novel educational opportunities \cite{Mollick_2023}, for example, by personalizing content or providing almost limitless varied examples.  

\begin{figure}
    \centering
    \includegraphics[width=\textwidth]{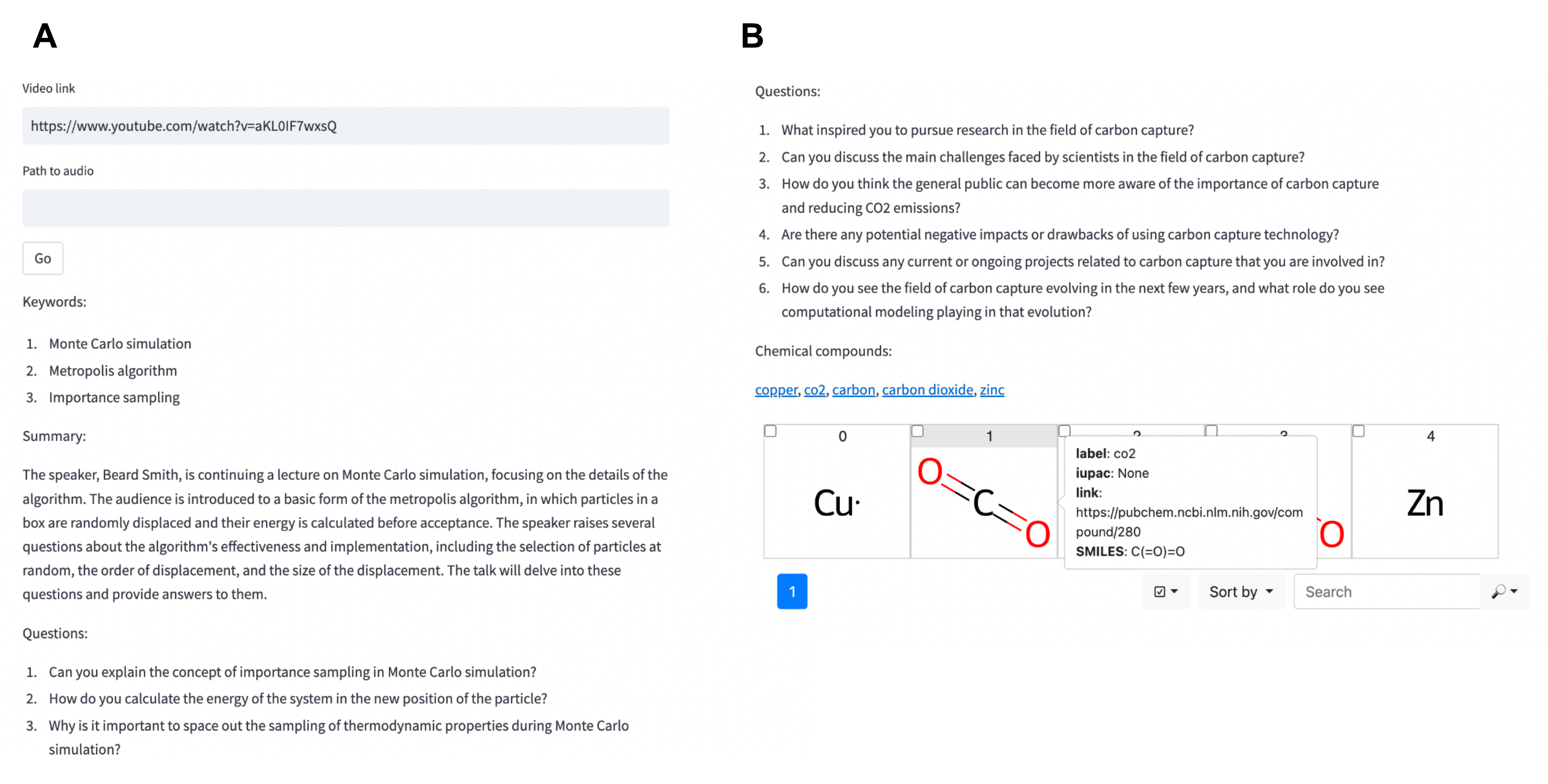}
    \caption{\emph{The I-digest interface.} A video (e.g., of a lecture recording) can be described using the Whisper model. Based on the transcript, an \gls{llm} can generate questions (and answers). Those can assist students in their learning. The \gls{llm} can also detect mentions of chemicals and link to further information about them (e.g., on PubChem \cite{pubchem,pugrest,PubChem2022}).}
    \label{fig:i-digest}
\end{figure}

The I-Digest (Information-Digestor) hackathon team (Beatriz  Mouriño, Elias Moubarak, Joren Van Herck, Sauradeep Majumdar, Xiaoqi Zhang) created a path toward such a new educational opportunity by providing students with a digital tutor based on course material such as lecture recordings. 
Using the Whisper model \cite{radford2022robust}, videos of lecture recordings can be transcribed to text transcripts. The transcripts can then be fed into an \gls{llm} with the prompt to come up with questions about the content presented in the video (\Cref{fig:i-digest}). 
In the future, these questions might be shown to students before a video starts, allowing them to skip parts they already know or after the video, guiding students to the relevant timestamps or additional material in case of an incorrect answer. 

Importantly, and in contrast to conventional educational materials, this approach can generate a practically infinite number of questions and could, in the future, be continuously be improved by student feedback. 
In addition, it is easy to envision extending this approach to consider lecture notes or books to guide the students further or even recommend specific exercises. 


\section{Conclusion}
The fact that the groups were able to present prototypes that could do quite complex tasks in such a short time illustrates the power of  \glspl{llm}. 
Some of these prototypes would have taken many months of programming just a few months ago, but the fact that \glspl{llm} could reduce this time to a few hours is one of the primary reasons for the success of our hackathon. 
Combined with the time-constrained environment in teams (with practically zero cost of \enquote{failure}), we found more energy and motivation. The teams delivered more results than in most other hackathons we participated in. 

Through the \gls{lift} framework, one can use \glspl{llm} to address problems that could already be addressed with conventional approaches---but in a much more accessible way (using the same approach for different problems), while also reusing established concepts such as $\Delta$-ML.
At the same time, however, we can use \glspl{llm} to model chemistry and materials science in novel ways; for example, by incorporating context information such as \enquote{fuzzy} design rules or directly operating on unstructured data. 
Overall, a common use case has been to use \Glspl{llm}  to deal with \enquote{fuzziness} in programming and tool development. 
We can already see tools like Copilot and ChatGPT being used to convert \enquote{fuzzy abstractions} or hard-to-define tasks into code. 
These advancements may soon allow everyone to write small apps or customize them to their needs (end-user programming). Additionally, we can observe an interesting trend in tool development: most of the logic in the showcased tools is written in English, not in Python or another programming language. 
The resulting code is shorter, easier to understand, and has fewer dependencies because \glspl{llm} are adept at handling fuzziness that is difficult to address with conventional code. This suggests that we may not need more formats or standards for interoperability; instead, we can simply describe existing solutions in natural language to make them interoperable. Exploring this avenue further is exciting, but it is equally important to recognize the limitations of \glspl{llm}, as they currently have limited interpretability and lack robustness.

It is interesting to note that none of the projects relied on the knowledge or understanding of chemistry by \glspl{llm}. Instead, they relied on general reasoning abilities and provided chemistry information through the context or fine-tuning. However, this also brings new and unique challenges. All projects used the models provided by OpenAI's \gls{api}. While these models are powerful, we cannot examine how they were built or have any guarantee of continued reliable access to them. 

Although there are open-source language models and techniques available, they are generally more difficult to use compared to simply using OpenAI's \gls{api}.  Furthermore, the performance of language models can be fragile, especially for zero- or few-shot applications.
To further investigate this, new benchmarks are needed that go beyond the tabular datasets we have been using for \gls{ml} for molecular and materials science---we simply have no frameworks to compare and evaluate predictive models that use context, unstructured data, or tools. Without automated tests, however, it is difficult to improve these systems systematically. 
On top of that, consistent benchmarking is hard because de-duplication is ill-defined even if the training data are known. 
To enable a scientific approach to the development and analysis of these systems, we will also need to revisit versioning frameworks to ensure reproducibility as systems that use external tools depend on the exact versions of training data, \gls{llm}, as well as of the external tools and prompting setup.

The diversity of the prototypes presented in this work shows that the potential applications are almost unlimited, and we can probably only see the tip of the iceberg---for instance, we didn't even touch modalities other than text thus far.

Given these new ways of working and thinking, combined with the rapid pace of developments in the field, we believe that we urgently need to rethink how we work and teach. 
We must discuss how we ensure safe use \cite{campbell2023censoring}, standards for evaluating and sharing those models, and robust and reliable deployments. But we also need to discuss how we ensure that the next generation of chemists and materials scientists are proficient and critical users of these tools---that can use them to work more efficiently while critically reflecting on the outputs of the systems. 
We believe that to truly leverage the power of \glspl{llm} in the molecular and material sciences, we need a community effort---including not only chemists and computer scientists but also lawyers, philosophers, and ethicists: the possibilities and challenges are too broad and profound to tackle alone.

\section*{Acknowledgements}

We would like to specifically thank Jim Warren (NIST) for his contributions to discussions leading up to the hackathon and his participation as a judge during the event. We would also like to thank Anthony Costa and Christian Dallago (NVIDIA) for supporting the hackathon.

\noindent
\textbf{B.B., I.T.F, and ZH} acknowledge support from the the National Science Foundation awards \#2226419 and \#2209892. This work was performed under the following financial assistance award 70NANB19H005 from the U.S. Department of Commerce, National Institute of Standards and Technology as part of the Center for Hierarchical Materials Design (CHiMaD). 

\noindent
\textbf{K.J.S, A.S.} acknowledge support from the the National Science Foundation award \#1931306.

\noindent
\textbf{K.M.J., S.M., J.v.H., X.Z., B.M., E.M., and B.S.}\ were supported by the MARVEL National Centre for Competence in Research funded by the Swiss National Science Foundation (grant agreement ID 51NF40-182892) and the USorb-DAC Project, which is funded by a grant from The Grantham Foundation for the Protection of the Environment to RMI’s climate tech accelerator program, Third Derivative. B.M. was further supported by the European Union’s Horizon 2020 research and innovation programme under the Marie Skłodowska-Curie grant agreement No. 945363.

\noindent
\textbf{M.C.R., S.C., and A.D.W.}\ were supported by the National Science Foundation and the National Institute of General Medical Sciences under Grant No. 1764415 and award number R35GM137966, respectively.

\noindent
\textbf{Q.A.}’s contribution to this work was supported by the National Center for Advancing Translational Sciences of the National Institutes of Health under award number U18TR004149. The content is solely the responsibility of the authors and does not necessarily represent the official views of the National Institutes of Health. 

\noindent
\textbf{M.V.G.}\ acknowledges support from the Spanish National Research Council (CSIC) through the Programme for internationalization i-LINK 2021 (Project LINKA20412), and from the Spanish Agencia Estatal de Investigación (AEI) through the Grant TED2021-131693B-I00 funded by MCIN/AEI/ 10.13039/501100011033 and by the “European Union NextGenerationEU/PRTR” and through the Ramón y Cajal Grant RYC-2017-21937 funded by MCIN/AEI/ 10.13039/501100011033 and by “ESF Investing in your future”.

\noindent
The \texttt{datalab} project (\textbf{M.L.E.}, \textbf{B.E.S.} and \textbf{J.D.B.}) has received funding from the European Union's Horizon 2020 research and innovation programme under grant agreement 957189 (DOI: 10.3030/957189), the Battery Interface Genome - Materials Acceleration Platform (BIG-MAP), as an external stakeholder project. \textbf{M.L.E.} additionally thanks the BEWARE scheme of the Wallonia-Brussels Federation for funding under the European Commission's Marie Curie-Skłodowska Action (COFUND 847587). \textbf{B.E.S.} acknowledges support from the UK's Engineering and Physical Sciences Research Council (ESPRC).

\noindent
\textbf{B.P.}\ acknowledges support from the National Science Foundation through NSF-CBET Grant No. 1917340. The authors thank Phung Cheng Fei, Hassan Harb, and Vinayak Bhat for their helpful comments on this project.

\noindent
\textbf{D.C. and L.C.B.}\ thank NSF DGE-2022040 for the aiM NRT funding support.

\noindent
\textbf{K.C.}\ thank the National Institute of Standards and Technology for funding, computational, and data-management
resources. Please note certain equipment, instruments, software, or materials are identified in this paper in order to specify the experimental procedure adequately. Such identification is not intended to imply recommendation or endorsement of any product or service by NIST, nor is it intended to imply that the materials or equipment identified are necessarily the best available for the purpose.

\noindent
\textbf{A.K.G., G.W.M., A.I., and W.A.d.J.} were supported by the U.S. Department of Energy, Office of Science, Basic Energy Sciences, Materials Sciences and Engineering Division under Contract No. DE-AC02-05CH11231, FWP No. DAC-LBL-Long, and by the U.S. Department of Energy, Office of Science, Office of High Energy Physics under Award Number DE-FOA-0002705.

\noindent
\textbf{M.B, B.R., and P.S.}\ were supported by the NCCR Catalysis (grant number 180544), a National Centre of Competence in Research funded by the Swiss National Science Foundation.

\noindent
\textbf{S.G.R. and J.L.}\ acknowledge the generous support of Eric and Wendy Schmidt, and the core funding of the Francis Crick Institute, which receives its funding from Cancer Research UK, the UK Medical Research Council, and the Wellcome Trust.


\bibliographystyle{achemso}
\bibliography{refs}



\section*{Supplementary material}

# Supporting Information for: 14 Examples of How LLMs Can Transform Materials Science and Chemistry: A Reflection on a Large Language Model Hackathon

Kevin Maik Jablonka 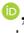,[1, *] Qianxiang Ai 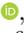,[2, †] Alexander Al-Feghali 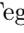,[3, †] Shruti Badhwar 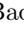,[4, †]
Joshua D. Bocarsly 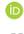,[5, †] Andres M Bran 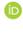,[6, 7, †] Stefan Bringuier 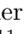,[8, †] L. Catherine Brinson 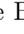,[9, †]
Kamal Choudhary 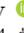,[10, †] Defne Circi 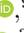,[9, †] Sam Cox 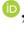,[11, †] Wibe A. de Jong 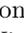,[12, †]
Matthew L. Evans 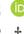,[13, 14, †] Nicolas Gastellu 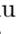,[3, †] Jerome Genzling 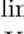,[3, †] María Victoria Gil 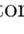,[15, †]
Ankur K. Gupta 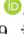,[12, †] Zhi Hong 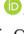,[16, †] Alishba Imran,[17, †] Sabine Kruschwitz 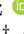,[18, †] Anne Labarre 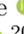,[3, †]
Jakub Lála 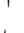,[19, †] Tao Liu 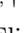,[3, †] Steven Ma 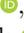,[3, †] Sauradeep Majumdar 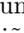,[1, †] Garrett W. Merz 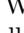,[20, †]
Nicolas Moitessier 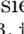,[3, †] Elias Moubarak 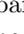,[1, †] Beatriz Mouriño 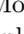,[1, †] Brenden Pelkie 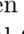,[21, †]
Michael Pieler 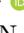,[22, 23, †] Mayk Caldas Ramos 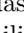,[11, †] Bojana Ranković 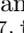,[6, 7, †] Samuel G. Rodriques 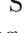,[19, †]
Jacob N. Sanders 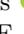,[24, †] Philippe Schwaller 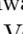,[6, 7, †] Marcus Schwarting,[25, †] Jiale Shi 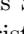,[2, †]
Berend Smit 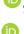,[1, †] Ben E. Smith 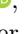,[5, †] Joren Van Herck 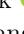,[1, †] Christoph Völker 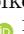,[18, †] Logan Ward 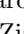,[26, †]
Sean Warren 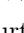,[3, †] Benjamin Weiser 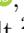,[3, †] Sylvester Zhang,[3, †] Xiaoqi Zhang 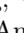,[1, †] Ghezal Ahmad Zia 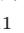,[18, †]
Aristana Scourtas 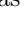,[27] KJ Schmidt,[27] Ian Foster 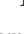,[28] Andrew D. White 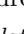,[11] and Ben Blaiszik 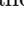[27, ‡]

[1] *Laboratory of Molecular Simulation (LSMO),*
*Institut des Sciences et Ingénierie Chimiques,*
*Ecole Polytechnique Fédérale de Lausanne (EPFL), Sion, Valais, Switzerland.*
[2] *Department of Chemical Engineering, Massachusetts Institute of Technology,*
*Cambridge, Massachusetts 02139, United States.*
[3] *Department of Chemistry, McGill University, Montreal, Quebec, Canada.*
[4] *Reincarnate Inc.*
[5] *Yusuf Hamied Department of Chemistry, University of Cambridge,*
*Lensfield Road, Cambridge, CB2 1EW, United Kingdom.*
[6] *Laboratory of Artificial Chemical Intelligence (LIAC),*
*Institut des Sciences et Ingénierie Chimiques,*
*Ecole Polytechnique Fédérale de Lausanne (EPFL), Lausanne, Switzerland.*
[7] *National Centre of Competence in Research (NCCR) Catalysis,*
*Ecole Polytechnique Fédérale de Lausanne (EPFL), Lausanne, Switzerland.*
[8] *Independent Researcher, San Diego, CA, United States.*
[9] *Mechanical Engineering and Materials Science, Duke University, United States.*
[10] *Material Measurement Laboratory, National Institute of Standards and Technology, Maryland, 20899, United States.*
[11] *Department of Chemical Engineering, University of Rochester, United States.*
[12] *Applied Mathematics and Computational Research Division,*
*Lawrence Berkeley National Laboratory, Berkeley, CA 94720, United States.*
[13] *Institut de la Matière Condensée et des Nanosciences (IMCN),*
*UCLouvain, Chemin des Étoiles 8, Louvain-la-Neuve, 1348, Belgium.*
[14] *Matgenix SRL, 185 Rue Armand Bury, 6534 Gozée, Belgium.*
[15] *Instituto de Ciencia y Tecnología del Carbono (INCAR),*
*CSIC, Francisco Pintado Fe 26, 33011 Oviedo, Spain.*
[16] *Department of Computer Science, University of Chicago, Chicago, Illinois 60637, United States.*
[17] *Computer Science, University of California,*
*Berkeley, Berkeley CA 94704, United States.*
[18] *Bundesanstalt für Materialforschung und -prüfung,*
*Unter den Eichen 87, 12205 Berlin, Germany.*
[19] *Francis Crick Institute, 1 Midland Rd, London NW1 1AT, United Kingdom.*
[20] *American Family Insurance Data Science Institute,*
*University of Wisconsin-Madison, Madison WI 53706, United States.*
[21] *Department of Chemical Engineering, University of Washington, Seattle, WA 98105, United States.*
[22] *OpenBioML.org*
[23] *Stability.AI*
[24] *Department of Chemistry and Biochemistry,*
*University of California, Los Angeles, CA 90095, United States.*
[25] *Department of Computer Science, University of Chicago, Chicago IL 60490, United States.*
[26] *Data Science and Learning Division, Argonne National Lab, United States.*
[27] *Globus, University of Chicago, Data Science and Learning Division, Argonne National Lab, United States.*

$^{28}$*Department of Computer Science, University of Chicago,*
*Data Science and Learning Division, Argonne National Lab, United States.*

**Contents**

———

\* mail@kjablonka.com
† These authors contributed equally
‡ blaiszik@uchicago.edu

# I. Predictive Modeling

## A. Leveraging LLMs for Accurate Molecular Energy Predictions

Table I. *LIFT for molecular atomization energies on the QM9-G4MP2 dataset.* Metrics for models tuned on 90% of the QM9-G4MP2 dataset (117,232 molecules), using 10% (13,026 molecules) as a holdout test set. Note that the metric used for the baseline results [1] is MAE, whereas this work used the MAD. The results indicate that the LIFT framework can also be used to build predictive models for atomization energies, that can reach chemical accuracy using a $\Delta$-ML scheme.

| mol. repr. & framework | G4(MP2) Atomization Energy | | (G4(MP2)-B3LYP) Atomization Energy | |
|---|---|---|---|---|
| | $R^2$ | MAD / eV | $R^2$ | MAD / eV |
| SMILES: GPTChem | 0.984 | 0.99 | 0.976 | 0.03 |
| SELFIES: GPTChem | 0.961 | 1.18 | 0.973 | 0.03 |
| SMILES: GPT2-LoRA | 0.931 | 2.03 | 0.910 | 0.06 |
| SELFIES: GPT2-LoRA | 0.959 | 1.93 | 0.915 | 0.06 |
| SchNet baseline | - | - | - | 0.0045 |
| FCHL baseline | - | 0.0223 | - | 0.0052 |

Accurate prediction of chemical properties has long been the ultimate objective in computational chemistry and materials science. However, the significant computational demands of precise methods often hinder their routine application in modeling chemical processes. The recent surge in machine learning development, along with the subsequent popularity of large language models (LLMs), offers innovative and effective approaches to overcome these computational limitations. Our project takes steps toward establishing a comprehensive, open-source framework that harnesses the full potential of LLMs to accurately model chemical problems and uncover novel solutions to chemical challenges. In this study, we assessed the capability of LLMs to predict the atomization energies of molecules at the G4(MP2) [2] level of theory from the QM9-G4MP2 dataset [3, 4] using solely string representations for molecules, specifically, SMILES [5] and SELFIES [6, 7]. G4(MP2) is a highly accurate composite quantum chemistry method, known for its accuracy within 1.0 kcal/mol for molecular energies compared to experimental values, making atomization energy an ideal property to predict to demonstrate the usefulness and impact of LLMs on the field of computational chemistry.

Jablonka et al. [8] recently demonstrated the potential of fine-tuning pre-trained LLMs on chemistry datasets for a broad array of predictive chemistry tasks. As an initial validation for our project, we fine-tuned generative pretrained transformer (GPT)-3 [9] to learn how to reproduce a molecule's atomization energy at the G4(MP2) level of theory, using its SMILES or SELFIES string through the prompt, "What is the G4MP2 atomization energy in kcal/mol of 'SMILES/SELFIES string of a molecule'?" Additionally, we fine-tuned LLMs to predict the atomization energy difference between B3LYP/6-31G(2df,p) and G4(MP2) levels of theory with the prompt, "What is the G4MP2 and B3LYP atomization energy difference in kcal/mol of 'SMILES/SELFIES string of a molecule'?", which mirrors the $\Delta$-machine learning ($\Delta$-ML) schemes [10] found in the existing literature.

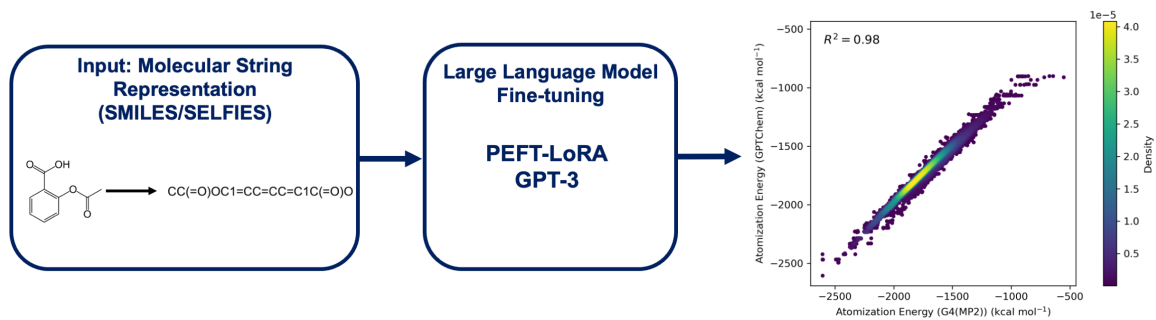

Figure 1. *Illustration of the molecular property prediction workflow, from molecule representation to model fine-tuning and performance evaluation.*

We fine-tuned the GPT-3 (Ada) model using 90% of the QM9-G4MP2 dataset (117,232 molecules) for eight epochs with the GPTChem [8] framework's default settings. The remaining 10 % (13,026 molecules) was kept as the hold-out set, following the same data split as Ward et al. [1], to evaluate the model's performance. Table I summarizes the regression metrics for the hold-out set. The strong correlation between the predicted and ground truth values suggests that the model effectively learned the structural information from the molecular string representation. Although the MAD remains relatively high compared to state-of-the-art models in the literature [1, 11] that utilize a molecule's full 3D structural information for descriptor construction, we achieved chemical accuracy ($< 1.0$ kcal/mol $\approx 0.04$ eV) for the $\Delta$-ML task. Consequently, this approach can predict G4(MP2) energies with high accuracy when B3LYP energies are available. We also compared the model's performance using SMILES and SELFIES molecular representations, with the former proving marginally superior for predicting atomization energies, possibly due to its more compact representation for molecules. We additionally calculated regression metrics for the G4MP2-Heavy dataset [1], the results of which are provided in Table II.

Table II. Regression metrics, (Coefficient of Determination), and MAD (Mean Absolute Deviation) for predicting G4(MP2) and (G4(MP2)B3LYP) atomization energies for the G4MP2-Heavy dataset using a fine-tuned GPT-3 model with GPTChem

| mol. repr. & framework | G4(MP2) Atomization Energy | | (G4(MP2)-B3LYP) Atomization Energy | |
|---|---|---|---|---|
| | $R^2$ | MAD / eV | $R^2$ | MAD / eV |
| SMILES: GPTChem | 0.583 | 6.02 | 0.856 | 0.13 |
| SELFIES: GPTChem | 0.146 | 9.44 | 0.659 | 0.15 |

While GPT-3 fine-tuning models are accessible through the OpenAI application programming interface (API), their usage costs can become prohibitive for larger datasets, rendering hyperparameter searches and other exploratory research economically unfeasible. Consequently, we aim to develop a free and open-source framework for fine-tuning LLMs to perform a wide range of predictive modeling tasks, encompassing chemical property prediction and inverse design.

To fine-tune a pre-trained LLM locally on a GPU instead of querying OpenAI's API, we employed the Hugging Face parameter efficient fine-tuning (PEFT) library [12] to implement the low-rank adaptors (LoRA) tuning paradigm [13]. Conventional fine-tuning updates all model parameters, utilizing pretrained weights from a large training dataset as a starting point for gradient descent. However, fine-tuning memory-intensive LLMs on consumer hardware is often impractical. The LoRA approach addresses this by freezing the model's weights and tuning a low-rank adapter layer rather than the entire model, parameterizing changes concerning the initial weights rather than the updated weights.

Using this approach, we fine-tuned the smallest version of GPT-2 [14] (124 million parameters) for 20 epochs on the same 90 % training set as used in GPTChem, allocating 10 % of that training set for validation, and computed metrics on the same 10 % hold-out set as in the GPTChem run, employing the same prompt structure. Although the model performs well, it demonstrates slightly inferior performance to GPT-3 on the G4MP2 task and moderately worse on the (G4(MP2)-B3LYP) task. This is not unexpected, given that GPT-3 is a more recent model with substantially more parameters than GPT-2 (175 billion vs. 124 million) and has exhibited superior few-shot performance on various tasks [15].

Moving forward, we plan to employ the LoRA tuning framework to fine-tune other models, such as LLaMA [16] and GPT-J, to investigate the impact of LLM selection on performance in chemistry-related tasks. Moreover, we intend to experiment with molecular-input representations beyond string formats to more accurately represent a molecule's 3D environment [17].

*One sentence summaries*

*a. Problem/Task* Predicting the atomization energies of molecules using large language models.

*b. Approach* Fine-tuning of GPT-3 `ada` model as well as PEFT of a small open-source model (GPT-2) on SMILES to either directly predict the atomization energies or the difference between a lower and a higher level of theory.

*c. Results and Impact*   Even though simpler, direct fine-tuning for a complicated property on SMILES leads to errors one order of magnitude higher than baselines, and the error can only be brought close to the baselines with an $\Delta - ML$ approach—first demonstration of $\Delta$-ML in the LIFT framework for chemistry.

*d. Challenges and Future Work*   Since the predictions without 3D coordinates is not satisfactory, a question for future work is how the approach would perform when provided with 3D coordinates.

## B. From Text to Cement: Developing Sustainable Concretes Using In-Context Learning

The inherently intricate chemistry and variability of feedstocks in the construction industry have limited the development of novel sustainable concretes to labor-intensive laboratory testing. This major bottleneck in material innovation has significant consequences due to the substantial contribution of $CO_2$ emissions of materials in use today. The production of Portland cement alone amounts to approximately 8 % of anthropogenic $CO_2$ emissions [18]. The increasing complexity of alternative raw materials and the uncertain future availability of established substitutes like fly ash and granulated blast furnace slag make the experimental development of more sustainable formulations time-consuming and challenging. Traditional trial-and-error approaches are ill-suited to efficiently explore the vast design space of potential formulations.

In previous studies, inverse design (ID) has been shown to accelerate the discovery of novel, sustainable, and high-performance materials by reducing labor-intensive laboratory testing [19–21]. Despite their potential, the adoption of these techniques has been impeded by several difficulties that are connected to the predictive model at the core of ID: Incorporating domain knowledge typically requires extensive data collection to accurately capture underlying relationships, which makes representing complex tasks in practice challenging due to the high costs of data acquisition. Furthermore, ID necessitates formulating research problems as search space vectors. This process can be unintuitive and challenging for lab personnel, limiting the comprehension and adoption of these techniques. Lastly, sparse training samples in high dimensions can lead to co-linearities and overfitting, negatively impacting prediction performance. With in-context learning

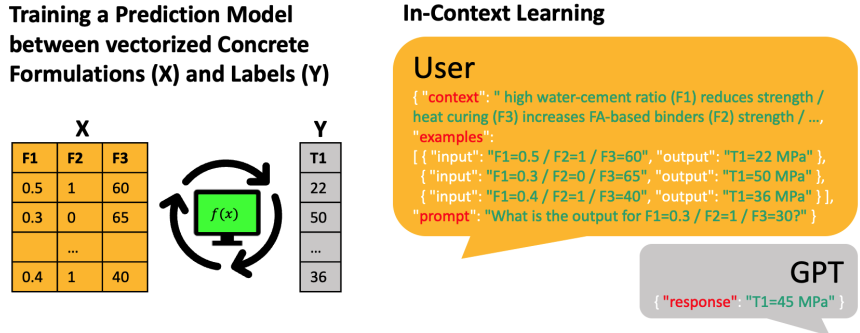

Figure 2. *Using LLMs to predict the compressive strength of concretes.* The left part illustrates the conventional approach for solving this task, i.e., training classical prediction models using tabular data. Using the LIFT framework LLM can also use tabular data but also leverage context information provided in natural language. Augmented with this context, in-context-learning with LLM leads to a performance that outperforms baselines such as RFs or GPRs.

(ICL), Jablonka et al. [8] and Ramos et al. [22] demonstrated that LLMs offer a solution by incorporating context and general knowledge, providing flexibility in handling non-numeric inputs and overcoming the limitations of traditional vector space formulations (Figure 2).

In this study, we have adopted an ICL approach based on a dataset from a study by Rao and Rao [23]. The dataset comprises 240 alternative and more sustainable concrete formulations based on fly ash and ground granulated slag binders, along with their respective compressive strengths. The goal is to compare the prediction performance of the compressive strength with ICL using the `text-davinci-003` model [24] against established methods, RF [25].

Randomly sampled training subsets containing ten formulations are drawn. The prediction performance is assessed on a separate, randomly sampled test set of 25 samples and evaluated using the coefficient of determination (R-squared) [26]. This process is repeated ten times to ensure more reliable results.

The experimental results reveal that ICL attains comparable performance to GPR but underperforms RF when provided with small training data sets (R-squared of 0.5, 0.54, and 0.67, respectively). However, when using general, qualitative concrete design knowledge, such as the influence of the water-to-cement ratio on strength, the models significantly reduce prediction outliers and ultimately surpass RF (R-squared = 0.71). When we incorrectly changed the context of the ratio of fly ash to GGBFS, it negatively affected the R-squared value for ICL, causing it to drop to 0.6. This misrepresentation of the rule led to a decrease in the model's predictive accuracy, demonstrating that the quality of the information included in the "fuzzy" context is critical to the overall performance of LLMs. It should be noted, however, that the impact on the R-squared value may vary depending on the importance of the rule in the overall context. That is, not all

changes in context have a similar impact, and the drop to 0.6 might occur only in the case of the ratio of fly ash to GGBFS. Other studies, such as those conducted in the LIFT work, [27] have shown LLM performance for minor changes in wording or the presence of noise in the features. In these experiments, the robustness of LIFT-based predictions was comparable to classical ML algorithms, making it a promising alternative for using fuzzy domain knowledge in predictive modeling.

LLMs have been shown to provide significant advantages in sustainable concrete development, including context incorporation, adaptable handling of non-numeric inputs, and efficient domain knowledge integration, surpassing traditional methods' limitations. ICLs simplifies formulating data-driven research questions, increasing accessibility and democratizing a data-driven approach within the building materials sector. This highlights LLMs potential to contribute to the construction industry's sustainability objectives and foster efficient solutions.

### *One sentence summaries*

*a. Problem/Task* Predicting the compressive strength of concrete formulations.

*b. Approach* ICL on language-interfaced tabular data, with and without "fuzzy" domain expertise (such as relationship between columns) provided in natural language.

*c. Results and Impact* Predictive models can be built without any training; if provided with domain expertise, those models outperform the baselines—first demonstration in chemistry of such fuzzy knowledge can be incorporated into models.

*d. Challenges and Future Work* ICL can be very sensitive to the prompt, hence future work should investigate the robustness of this approach.

## C. Molecule Discovery by Context

The escalating climate crisis necessitates the deployment of clean, sustainable fuels to reduce carbon emissions. Hydrogen, with its potential to prevent approximately 60 gigatons of $CO_2$ emissions by 2050, according to the World Economic Forum, stands as a promising solution [28]. However, its storage and shipping remain formidable challenges due to the necessity for high-pressure tanks. To address this, we sought new molecules to which hydrogen could be conveniently added for storage. Traditional screening methods, like brainstorming, are insufficient due to their limited throughput. This research proposes a novel method of leveraging ScholarBERT, [29] a pre-trained science-focused LLM, to screen potential hydrogen carrier molecules efficiently. This approach utilizes ScholarBERT's ability to understand and relate the context of scientific literature. The data used for this study consisted of three datasets. The "Known" dataset comprised 78 known hydrogen carrier molecules. The "Relevant" dataset included 577 molecules, all of which are structurally similar to the "Known" molecules. The "Random" dataset contained 111 randomly selected molecules from the PubChem database [30]. The first step involved searching for contexts for molecules in the Public Resource Dataset (PRD), which includes 75M English language research articles. These contexts (i.e. sentences that mentioned the molecule name) were then fed into ScholarBERT. For each context, three calculations were made:

1. the average of the last four encoder layers in ScholarBERT

2. the average embedding of all tokens constituting the molecule name as one contextualized embedding for this molecule, and

3. the average of all contextualized embeddings for a molecule as ScholarBERT's representation of this molecule.

Subsequently, we calculated the similarity between the known and candidate molecules. The definition of "similarity" used in this study was the cosine similarity between the ScholarBERT representations of two molecules. We then sorted the candidates based on the similarity score in descending order, with a higher score indicating greater potential as a hydrogen carrier. Figure 3 and 4 show the candidate molecules with the highest similarity to the known molecules. We can see that ScholarBERT does a passable job finding similar molecules from the random set. We do see that it favors finding molecules with 5- and 6-member rings, though with features we didn't expect, like halogens. On the other hand, ScholarBERT does a much better job when we reduce the search space to those with structural similarity. We see that molecules with 5-member rings, for instance, are found to be similar structurally and in how they are described in the literature via ScholarBERT.

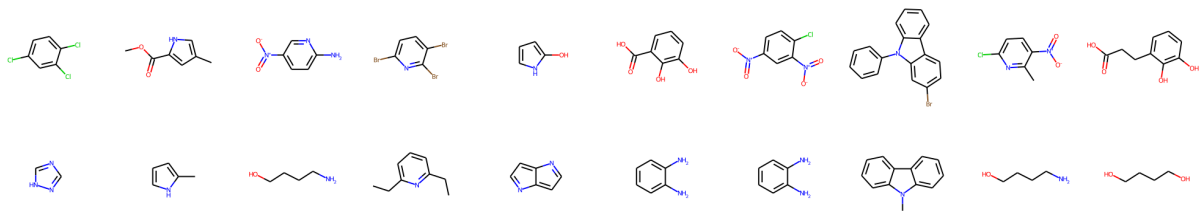

Figure 3. Each column shows a Known molecule on the bottom and its top candidate molecule from the Random set on the top

Based on our empirical data, computing the energy capacity (wt%$H_2$) and energy penalty (kJ/mol/$H_2$) of adding and removing $H_2$ to the molecule (which are the quantitative "success metrics" for this project) of a candidate molecule using traditional quantum chemistry takes around 30 seconds per molecule on a 64-core Intel Xeon Phi 7230 processor, whereas the proposed LLM approach can screen around 100 molecules per second on a V100 GPU, achieving a 3000 times speedup.

*One sentence summaries*

a. *Problem/Task*   Recommending hydrogen carrier molecules.

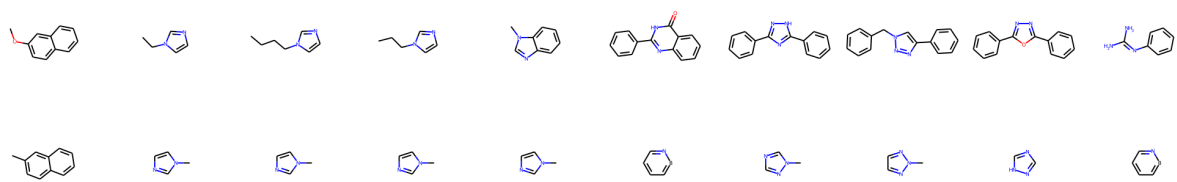

Figure 4. Each column shows a Known molecule on the bottom and its top candidate molecule from the Relevant set on the top

*b. Approach*   Ranking of molecules based on similarity of embeddings created with a language model by showing aggregating the embeddings of mentions of the molecules in scientific texts.

*c. Results and Impact*   Approach can recommend molecules with a success rate better than random.

*d. Challenges and Future Work*   More benchmarks compared to conventional generative modeling are needed.

### D. Text template paraphrasing with LLMs

#### 1. Problem

Text data is much trickier to augment for machine learning applications due to the discrete nature of the data modality. There are some traditional augmentation approaches for these tasks. However, they can be inefficient or still need extensive manual checks to be sure they deliver the desired results, especially for scientific or chemistry applications.

#### 2. Solution

To automate high-quality text data augmentations, LLMs have been explored by Dai and his coworkers [31] as a very recent and promising solution to this problem. We investigated such a setup in the scope of the OpenBioML chemistry NLP project (`https://github.com/OpenBioML/chemnlp`) to paraphrase text templates for the insertion of chemical raw data into natural language for LIFT. [27]

An example prompt is shown below. The outlined prompt setup has after "Question:" the desired task with additional information and after "Sentence:" the starting text template for the paraphrasing. The "Question:" and "Answer:" headers are not used if the LLM interface uses a chat interface, i.e., with OpenAI GPT-4.

---

**Example Prompt**

Question: Please paraphrase the sentence below ten times without changing the original meaning and the placeholder in the curly {} brackets. Please use all the placeholders in the curly {} brackets for every rephrased sentence.

Sentence: The molecule with the {SMILES_description} representation {SMILES_data} has a {target_name} of {target_value}{target_units}.

Answer:

---

**Example Response by GPT-4**

1. The {SMILES_description} representation of the molecule is {SMILES_data}, and its {target_name} is {target_value}{target_units}.

2. With a {SMILES_description} representation of {SMILES_data}, the molecule has a {target_name} of {target_value}{target_units}.

3. The molecule, represented by {SMILES_description} as {SMILES_data}, possesses a {target_name} of {target_value}{target_units}.

4. A molecule having a {SMILES_description} representation of {SMILES_data} also has a {target_name} equal to {target_value}{target_units}.

5. The {SMILES_description}-represented molecule with {SMILES_data} exhibits a {target_name} of {target_value}{target_units}.

6. For the molecule with a {SMILES_description} representation of {SMILES_data}, its {target_name} measures {target_value}{target_units}.

7. The {target_name} of the molecule having a {SMILES_description} representation of {SMILES_data} is {target_value}{target_units}.

8. Represented by {SMILES_description} as {SMILES_data}, the molecule's {target_name} is {target_value}{target_units}.

9. The molecule's {target_name} is {target_value}{target_units} when its {SMILES_description} representation is {SMILES_data}.

10. A {target_name} of {target_value}{target_units} is associated with the molecule whose {SMILES_description} representation is {SMILES_data}.

**Example Response by Anthropic Claude v1**

Here are 10 paraphrased sentences of the original sentence with the placeholders retained:

1. The molecule represented by the {SMILES_description} code {SMILES_data} has a {target_name} of {target_value}{target_units}.

2. The {SMILES_description} notation {SMILES_data} denotes a molecule with a {target_name} equal to {target_value}{target_units}.

3. A {target_name} of {target_value}{target_units} is possessed by the molecule depicted using {SMILES_description} as {SMILES_data}.

4. {SMILES_data} signifies the {SMILES_description} of the molecule which has a {target_name} of {target_value}{target_units}.

5. The {SMILES_description} {SMILES_data} characterizes the molecule that exhibits a {target_name} of {target_value}{target_units}.

6. A {target_name} equal to {target_value}{target_units} belongs to the molecule expressed with {SMILES_description} code {SMILES_data}.

7. {SMILES_data} denotes the {SMILES_description} for the molecule possessing a {target_name} of {target_value}{target_units}.

8. The molecule defined by {SMILES_description} {SMILES_data} presents a {target_name} of {target_value}{target_units}.

9. {SMILES_description} {SMILES_data} specifies the molecule which displays a {target_name} of {target_value}{target_units}.

10. A {target_name} amounting to {target_value}{target_units} is a property of the molecule represented with {SMILES_description} {SMILES_data}.

**Example Response by GPT-J**

The molecule with the {SMILES_description} representation {SMILES_A} has a {target_name} of {target_value}{target_units}. {SMILES_A} has a {target_name} of {target_units}. The molecule with the {target_units} representation {target_name} has a TARGET_UNITS of {target_value}{target_units}. {target_name} of {target_units} have a TARGET_UNITS of {target_value}{target_units}.

In the above answer, there is the {SMILES_description} representation of {SMILES_query}, but we don't use it in the sentence yet. And there is no curly brackets for the excepted answer.

*3. Impact*

The outlined approach allows to automatically create new paraphrased high-quality prompts for LIFT LLM training data very efficiently. With the additional paraphrased text templates, overfitting to special text passages should be avoided. We explore this setup in follow-up work in more detail.

*4. Lessons learned*

The outlined paraphrasing setup works well for the latest state-of-the-art models, e.g., OpenAI's GPT-4 and Anthropic's Claude v1. Less capable open-source models seem to lack the understanding of this paraphrasing task. Still, new and upcoming open-source LLM efforts could change that soon, enabling a cost-effective and broader application of this setup.

*One sentence summaries*

*a. Problem/Task*   Generation of many text-templates for language-interfaced fine-tuning of LLMs

*b. Approach*   Prompting of LLM to rephrase templates (with template syntax similar to Jinja).

*c. Results and Impact*   Large models (GPT-4, Claude), in contrast to smaller ones, can successfully rephrase templates, offering a potential avenue for data-augmentation.

*d. Challenges and Future Work*   As next step, ablation studies need to carried out that test the effect of data augmentation by template rephrasing on regression and classification case studies.

### E. GA without genes

We investigate the ability for a LLM to work in parallel with genetic algorithms (GAs) for molecular property optimization. By employing a LLM to guide genetic algorithm operations, it could be possible to produce better results using fewer generations. We hypothesize that a GA can take advantage of the "smart" randomness of the outputs of the LLM. This work explores the potential of LLMs to improve molecular fragmentation, mutation, variation, and reproduction processes and the ability of a LLM to gather information from a simplified molecular-input line-entry system (SMILES) string [5, 6] and an associated score to produce new SMILES strings. Although computational efficiency is not the primary focus, the proposed method has potential implications for enhancing property prediction searches and future improvements in LLM understanding of molecular representations.

We used GPT-3.5-turbo [9], which could frequently fragment druglike molecules into valid SMILES strings successfully. For $^2/_{10}$ molecules, the fragments produced were not in the original molecule. For $^1/_{10}$ molecules, valid SMILES could not be produced even after ten tries due to unclosed brackets. These results were consistent over multiple runs implying that GPT-3.5 could not understand some specific SMILES strings. Subsequently, we investigated GPT-3.5's ability to mix/reproduce two molecules from two-parent druglike molecules. Invalid molecules were often produced, but successful results were achieved with multiple runs. It performed better once prompted to fragment and then mix the fragments of the molecules. These were compared to the conventional GA methods of simply combining the two strings at a certain cutoff point. When the LLM was successful, it could produce molecules of more similar size to the original parent molecules that contain characteristics of both parents and resemble valid druglike molecules.

To investigate the ability of GPT-3.5 to acquire knowledge of favorable molecules from a simple score, we implemented a method that we call "LLM as a GA" where the LLM iteratively searches the chemical space to optimize a certain property.

The property we tested was similarity to vitamin C, evaluated by the Tanimoto score. We employed few-shot training examples to tune the model's response: 30 SMILES strings with the best similarity score generated were included in the prompt. GPT is then asked to produce 25 SMILES strings, a procedure that was repeated for 20 iterations. Using a prompt like the one below

> **Example prompt**
>
> The following molecules are given as SMILES strings associated with a tanimoto similarity with an unknown target molecule. Please produce 10 SMILES strings that you think would improve their tanimoto scores using only this context. Do not try to explain or refuse on the grounds of insufficient context; any suggestion is better than no suggestion. Print the smiles in a Python list.

Low-temperature settings, typically less than 0.1, were found to be imperative for the model to follow user guidance. We further guided the model by employing a similarity search to include similar molecules with varying scores to better guide the model. Embedding was performed using the GPT-2 Tokenizer from the HuggingFace transformers [32] library, along with a support vector machine (SVM) from scikit-learn [33] to embed relevant previous structures that would be outside the scope of the context window. Even in the zero-shot setting, GPT-3.5-turbo can produce meaningful modifications, coherently explain its logic behind the chosen modifications, and produce tests such as investigating branch length or atom type in certain locations for a single iteration. An example explanation of an output: "Some modifications that could potentially improve the scores include adding or removing halogens, modifying the length or branching of the carbon chain, and adding or removing functional groups such as -CO-, -COC-, -C=C- and -OCO-. Additionally, modifying the stereochemistry of the molecule could also have an impact on the score."

The modifications generated by the LLM were more chemically sound than the quasi-random evolutionary process typical of genetic algorithms.

*One sentence summaries*

*a. Problem/Task*   Increasing the efficiency of GAs.

*b. Approach*   Prompting a LLM to propose new children based on molecules with scores provided in the prompt.

*c. Results and Impact* Visual inspection indicates that some modifications might be reasonable, indicating a potential for more efficient genetic operations using LLMs.

*d. Challenges and Future Work* More systematic investigations on the performance and robustness compared to conventional GA operations are needed.

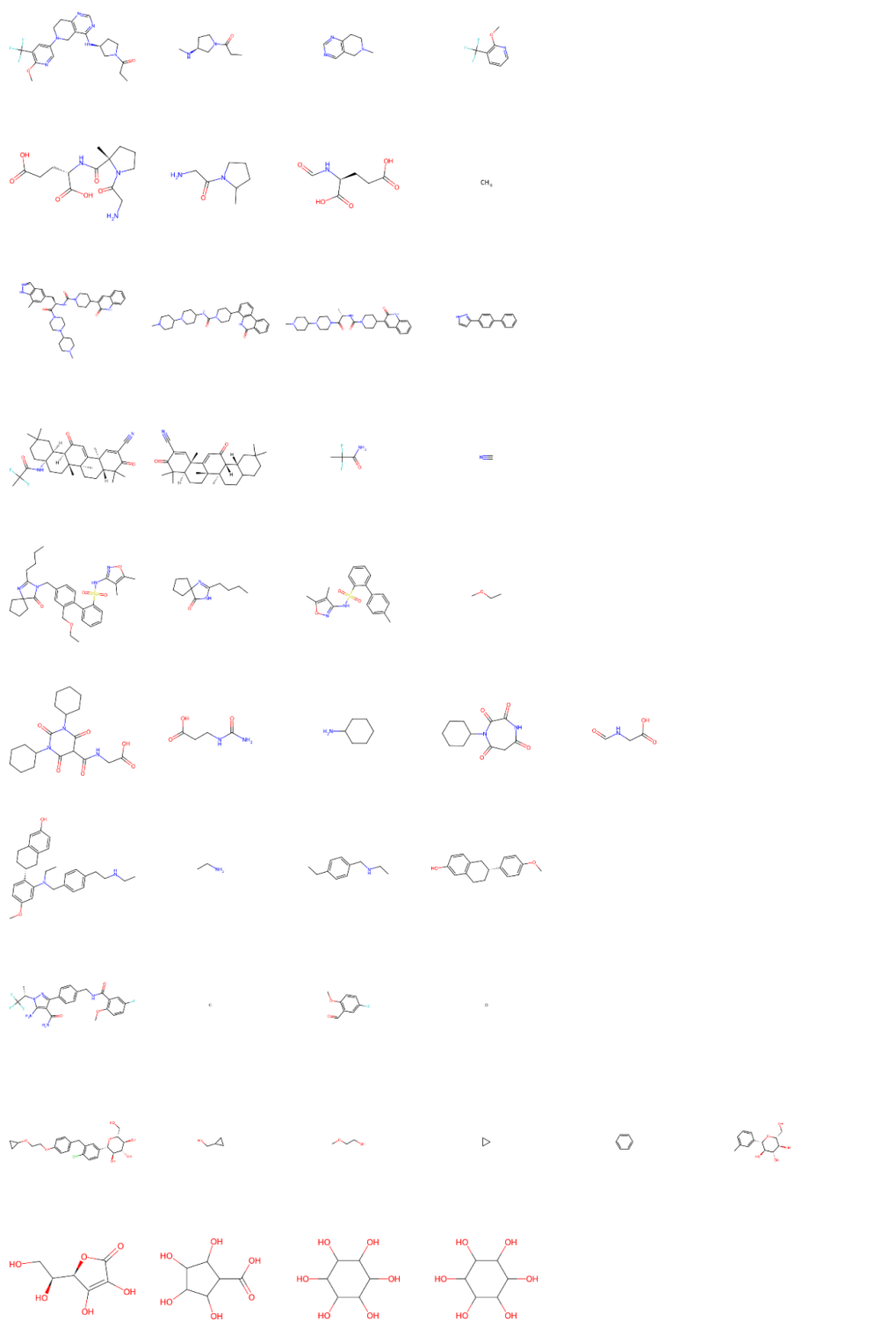

Figure 5. Using GPT to fragment molecules. Original molecules are in column one with LLM created fragment to the right. The LLM can frequently fragment molecules into valid SMILES strings successfully. $^2/_{10}$ times fragments produced were not in the original molecule (rows 6 and 10). For $^1/_{10}$ molecules, valid SMILES were able to be produced even after ten attempts (row 8)

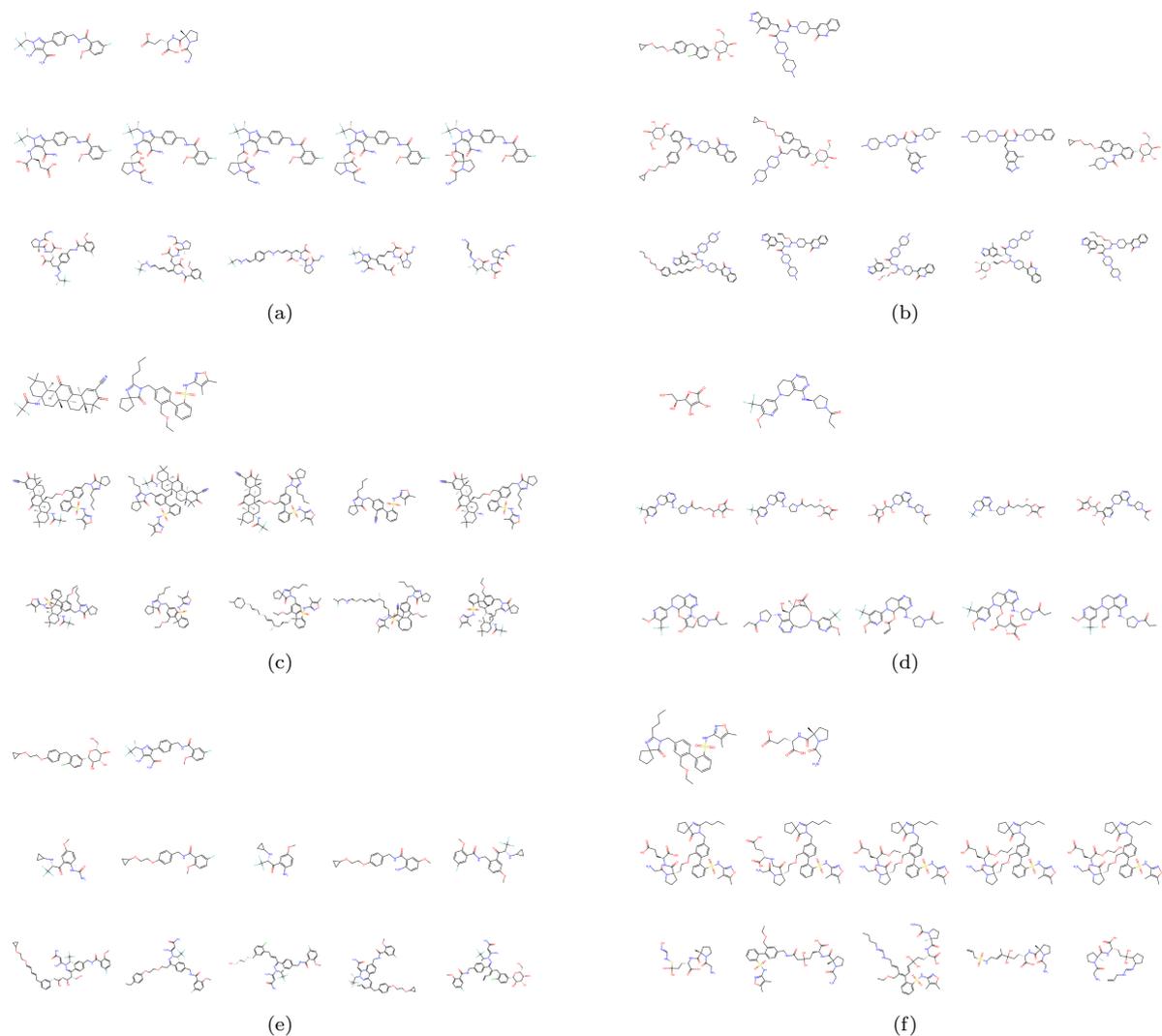

(a)

(b)

(c)

(d)

(e)

(f)

Figure 6. Using GPT-3.5-turbo to reproduce/mix molecules. Two original parent molecules on 1st row, followed by LLM created children, followed by conventional GA string splicing children for comparison

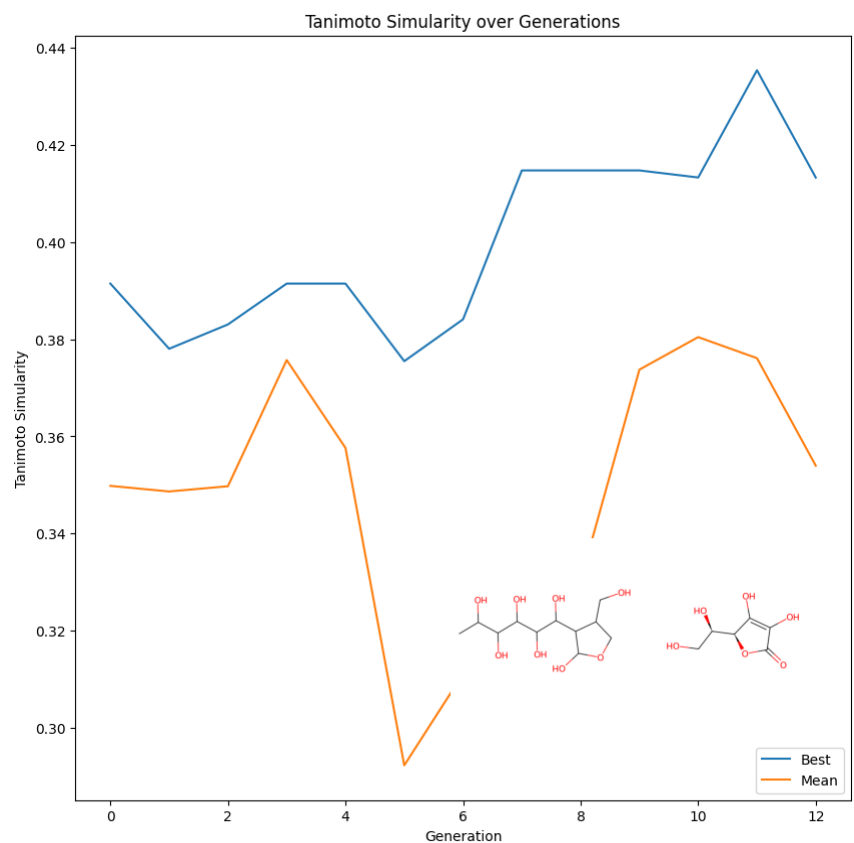

Figure 7. Tanimoto similarity to vitamin C as a function of GA generations. Conventional GA run for 30 generations and the best score (most similar to vitamin C) of each generation is given to the LLM as a LLM along with its associated Tanimoto similarity score to Vitamin C. LLM was then asked to create new molecules and improve the score for 12 generations. Multiple new best molecules were found using LLM as shown by the blue line.

## II. Automation and novel interfaces

### A. Using chain-of-thought and chemical tools to answer materials questions

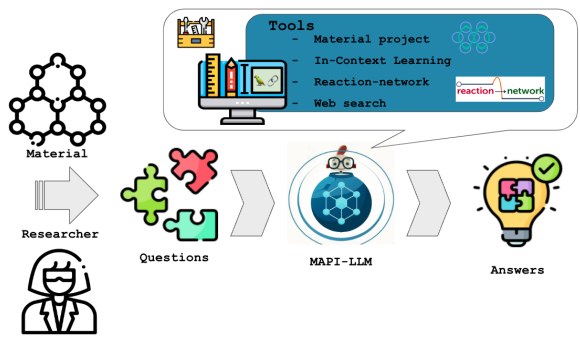

Figure 8. *Schematic overview of the MAPI-LLM workflow.* It uses LLMs to process the user's input and decide which available tools (e.g., Materials Project API, and Google Search) to use following an iterative chain-of-thought procedure. In this way, it can answer questions such as "Is the material AnByCz stable?".

LLMs have demonstrated remarkable success in various tasks [34–36]. Recently, LLMs have gained attention in chemistry, demonstrating exceptional ability to model chemical systems [37] and predicting tabular data [8, 22, 27]. Predicting the properties of materials is challenging since it requires computationally intensive techniques, such as density functional theory (DFT) [38–40]. Data-driven models offer a viable option to balance accuracy and computational time. Here we presented the MAPI-LLM, a multi-task package that employs LangChain [41] agents with access to multiple tools to address users' questions about materials.

It has been shown that providing chemistry-specific tools to an LLM allows the LLM to solve chemistry problems with significantly higher accuracy [42]. In a similar manner, we developed tools to iteratively query the Materials Project (MAPI) dataset [43] and utilize the reaction-network package [44], among others. MAPI-LLM can process user prompts in natural language using LLMs and follow a chain of thought (COT) [45] approach to determine the most suitable tools and inputs to answer the prompt. Due to MAPI-LLM's design, more tools can be added as needed, and tools can be combined (multiple tools can be used for a given prompt), opening the door for a large variety of applications. Figure 8 illustrates MAPI-LLM's capabilities. The code for the app is available in `https://github.com/maykcaldas/MAPI_LLM`, and a graphical user interface (GUI) is implemented in `https://huggingface.co/spaces/maykcaldas/MAPI_LLM`.

An important feature implemented into MAPI-LLM is a technique known as ICL [9], which allows the model to learn from the context within the prompt. For example, users can use MAPI-LLM's tool to query the MAPI dataset, first triggering the dataset search in the COT. However, if the desired material is not found in the dataset, MAPI-LLM still has access to other tools (such as ICL) to build context around the user prompt and adjust the COT actions to make a prediction. Another interesting tool is the ability to use the reaction-network package [44], which is a package for predicting inorganic reaction pathways. We showed the promising capabilities of MAPI-LLM by simply asking for reactions that use a given material as reactants or products. It can suggest such reactions for material synthesis or decomposition.

We built from the knowledge that LLMs are suitable for such tasks of interest in this application, for instance, classification and regression tasks [8]. Nevertheless, this application still needs a systematic validation of its predictions, such as the reinforcement learning from human feedback (RLHF) implementation in GPT-3.5 [46].

*One sentence summaries*

*a. Problem/Task* Answering complex materials science questions based on reliable data and tools.

*b. Approach* LLM-based agent in the ReAct framework that has access to tools such as the Materials Project API and uses ICL to answer questions for materials that are not in the materials project.

*c. Results and Impact* Coupling of tools allows answering questions that none of the tools or LLMs alone could solve by themselves, providing a very accessible interface to materials informatics tools.

*d. Challenges and Future Work* If a description of tools is incorporated in the prompt, this limits the number of tools that can be coupled. In addition, LLM agents still tend to not perform equally well on all prompts, and systematic investigation to better understand this and to increase the robustness is needed.

## B. sMolTalk

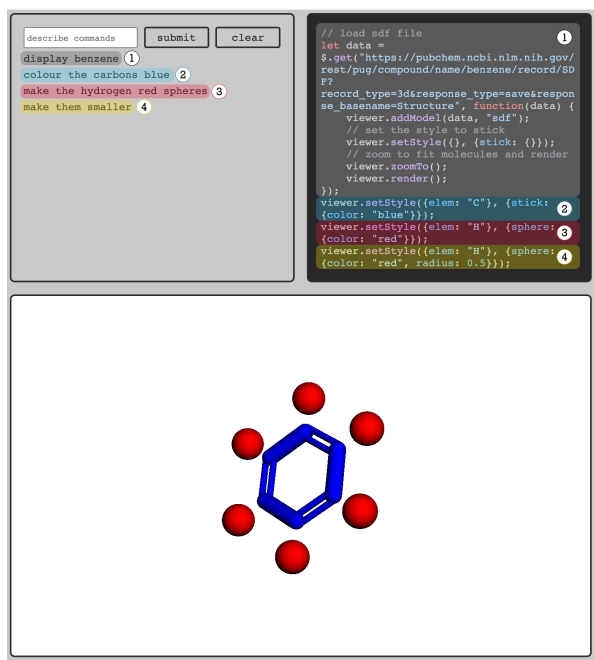

Figure 9. *The sMolTalk interface.* Based on few-shot prompting LLMs can create code for visualization tools such as `3dmol.js`.

Since the advent of 3D visualization methods, chemists have employed computers to display their molecules of interest to better understand their underlying structure and properties. Nevertheless, a lot of chemists are not equipped with the required coding skills to use and customize their visualizations. Depending on the package, and its associated documentation, chemists might end up spending hours to days learning the details of the specific visualization software.

We developed a natural language interface that generates code for `3dmol.js`, an open-source visualization JavaScript library [47], meaning the visualizations are run in a web browser (Figure 9). The user input is fed into ChatGPT API, using the GPT-3.5-turbo model. We use in-context learning (few-shot prompting), giving several examples of the user input with the expected JavaScript code that manipulates the `3dmol.js` viewer. Before the user submits further commands, we update the prompt with the current state of the viewer.

The current implementation might lead to a one-stop solution for visualizing and retrieving properties for molecules. This would accelerate chemists' workflow for querying information about molecules. Furthermore, if an LLM is able to control structural software, it might be possible to perform reasoning on the molecular structure itself. For instance, in drug discovery, one may ask what functional group of the ligand needs to be changed for binding affinity to the protein to increase. Another example might involve proteins, looking at what amino acid residues could be mutated to cysteines in order to create new disulfide bonds between chains. This would presumably require specific fine-tuning and equipping the LLM with more tools. The approach of generating code and structural reasoning might be similar but is most likely going to require a different set of tools that were specifically developed for protein structure manipulation (such as `PyMoL` [48], or `MolStar` [49]). Then, another set of highly accurate tools for binding affinity predictions or protein folding is also required. The major problem encountered is prompt leakage, where examples from in-context learning would leak into the actual LLM output. For the best evaluation, it is best to have as few and as different examples as possible. Moreover, although OpenAI's GPT models can sometimes correctly recall protein data bank (PDB) IDs of proteins or Chemical Abstract Services (CAS) numbers of compounds, it's not reliable, making tooling the models with API calls to PubChem, or the PDB, much more robust. We are currently developing an agent based on the ReAct approach [50] tooled with these APIs so that correct structures are always retrieved (i.e., to avoid the LLM needs to remember internally all such IDs). This framework would

then help us iteratively add tools to the agent, creating a chatbot one can query about any molecule of interest, including the structural reasoning task mentioned above. Lastly, we hypothesize we could improve the generation of `3dmol.js` code by using self-instruct fine-tuning. Using an external LLM with access to the documentation would create a dataset that could be used for fine-tuning. The same approach might be utilized for generating code for any other type of software, not just visualization packages. Therefore, such LLM could control molecular dynamics software, such as `LAMMPS` [51], or `GROMACS` [52].

### One sentence summaries

*a. Problem/Task*  Making bioinformatics tools, in particular the visualization software `3dmol.js` accessible to non-experts.

*b. Approach*  Chat-interface by prompting a LLM to produce commands to `3dmol.js`, which are then passed to the software.

*c. Results and Impact*  The LLM can, without consulting the documentation, generate code that often successfully performs the requested actions, demonstrating that LLM might help make tools more accessible by providing access to complex interfaces via natural language.

*d. Challenges and Future Work*  The approach suffers from robustness issues, which might be overcome with conditioning on a knowledge base, more advanced decoding strategies, or error handling (e.g., giving the agent access to error messages and logs and prompting it to reflect).

## C. `whinchat`: A Conversational ELN Interface

Laboratory data management systems such as ELNs and laboratory information systems (LIMS) can record rich data about experimental samples and the experiments that have been performed on them. Then, the user interfaces for these systems typically allow researchers to explore their samples with various pre-programmed visualizations. However, it is difficult for a traditional user interface to have enough flexibility to capture the richness and diversity of real, interconnected, experimental data. Here, our aim was to investigate whether an LLM could process digital laboratory data and serve as a flexible, conversational interface to a data management system. As a proof-of-concept for this approach, we introduced an automated data management assistant into the open-sourced "Datalab" software, which we have been developing to address needs in experimental materials chemistry (`https://github.com/the-grey-group/datalab`). Datalab can be used to capture data and metadata about materials chemistry samples, their synthesis constituents, and resulting devices. The system tracks each sample's provenance through the research lifecycle and makes this data available via a JavaScript object notation (JSON) API that can be explored programmatically or with a Vue.js-based web user interface (UI). During the hackathon, we embedded a prompted LLM into the UI and gave it access to raw responses from datalab's JSON API. We used the most recent publicly available models from OpenAI (gpt-3.5-turbo) via their Python API. We found that the LLM does a very good job at interpreting JSON and inferring the meaning of fields (including linked data) and can:

1. *Contextualize existing data:* Explain a set of related experiments from linked JSON responses, resolve acronyms/ short-hand used by experimentalists or create simple text summaries.

   > **Contextualizing existing data**
   >
   > **Assistant** Hello! I am your virtual assistant and I am here to help you manage your experimental data. How can I assist you today?
   >
   > **User** Can you provide a brief summary of this experiment?
   >
   > **Assistant** Based on the provided JSON data, this experiment involves the synthesis and characterization of $NaCoO_2$ as a battery cathode material. The synthesis method used was solid-state reaction, and the starting materials used were sodium carbonate and cobalt (II,III) oxide. The synthesized $NaCoO_2$ was then used to make an electrode, which was characterized using X-ray diffraction. The electrode was then incorporated into a coin cell battery, which was cycled electrochemically and characterized using various techniques.

2. *Generate text and graphical* representations of the data: Aggregate data from different fields in the JSON response and convert to formats such as tables (potentially formatted for a particular journal), CSV, and even mermaid.js graph visualizations.

3. *Suggest future experiments:* The assistant readily identifies related materials to study, additives to try, additional characterization techniques, and synthesis protocols.

   > **Suggest experiments**
   >
   > **User:** What NMR nuclei could I probe in this sample?
   >
   > **Assistant:** Based on the information provided in the JSON file, the sample is $NaCoO_2$, which contains sodium, cobalt, and oxygen. Therefore, the NMR nuclei that could be probed in this sample are:
   >
   > (a) Sodium-23 (23Na) NMR: This can provide information about the local environment and coordination of sodium ions in the $NaCoO_2$ lattice.
   >
   > (b) Cobalt-59 (59Co) NMR: This can provide information about the local environment and coordination of cobalt ions in the $NaCoO_2$ lattice.
   >
   > (c) Oxygen-17 (17O) NMR: This can provide information about the local environment and coordination of oxygen ions in the $NaCoO_2$ lattice.

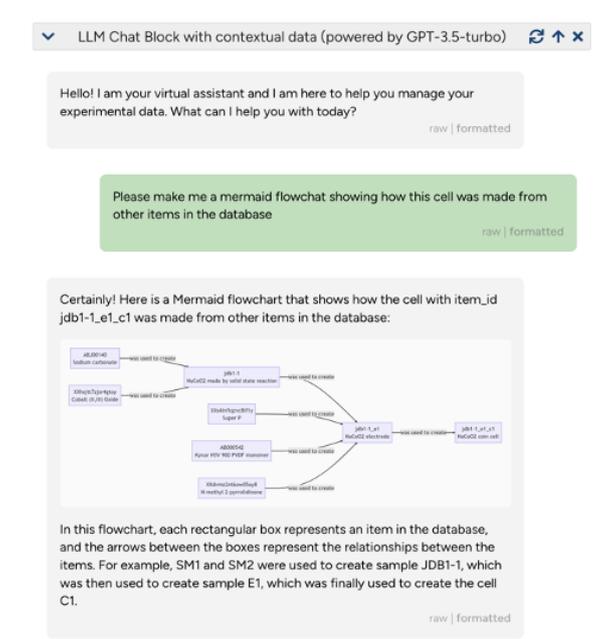

Figure 10. The chatbot can generate mermaid.js markdown code, which our interface automatically recognizes and translates into the visualization.

It is worth noting that NMR spectroscopy is a powerful technique for probing the local environment of atoms in a material, but it requires a relatively large amount of sample and may not be sensitive to all types of nuclei. Therefore, it may be necessary to use other techniques in conjunction with NMR to fully characterize the sample.

**User:** What techniques (that I haven't already used) should I use in conjunction with the NMR?

**Assistant:** In addition to the techniques I mentioned earlier, you could use the following techniques in conjunction with NMR to fully characterize the $NaCoO_2$ sample:

(a) X-ray photoelectron spectroscopy (XPS): This can provide information about the chemical composition and oxidation state of the $NaCoO_2$ sample, which can help to interpret the NMR spectra.

(b) Transmission electron microscopy (TEM): This can provide information about the crystal structure, morphology, and defects of the $NaCoO_2$ sample, which can help to interpret the NMR spectra.

(c) Electrochemical impedance spectroscopy (EIS): This can provide information about the electrical properties and performance of the $NaCoO_2$ sample as an electrode material, which can help to interpret the NMR spectra.

(d) Thermogravimetric analysis (TGA): This can provide information about the thermal stability and decomposition behavior of the $NaCoO_2$ sample, which can help to interpret the NMR spectra.

By using a combination of these techniques, you can gain a more complete understanding of the $NaCoO_2$ sample

The assistant performs quite well at certain tasks out-of-the-box, but we believe it should also be possible to further increase its general utility by fine-tuning the model on the domain knowledge present in a research

group's Wiki and papers. Furthermore, while we have only given the model data from a single project so far (up to 8 samples/starting materials, 1700 tokens of JSON), it would be of great interest to provide the model with a larger context across multiple projects to attempt to facilitate cross-fertilization of ideas. One notable challenge in this area is the limited context size of currently available LLM models (e.g., 4097 tokens for GPT-3.5-turbo). Therefore, future work will investigate larger models (e.g., GPT-4 with 30K token context), as well as approaches to give existing LLMs access to larger context (e.g., an embedding-based approach or allowing an LLM agent to query the OpenAPI directly as needed). At present, we note that the scientific usefulness of this assistant is highly task- and model-dependent; however, any additional interface that can lower the barrier to improving data capture and dissemination in the field should be investigated further and will be a future development target for Datalab.

*One sentence summaries*

*a. Problem/Task* Providing very flexible access to data in ELNs/LIMS.

*b. Approach* Prompting of a large language model with questions provided in a chat interface and context coming from the response of the API of an LLM.

*c. Results and Impact* The system can successfully provide a novel interface to the data and let user interact with it in a very flexible and personalized way, e.g, creating custom summaries or visuals for which the developers did not implement specific tools.

*d. Challenges and Future Work* Since the current approach relies on incorporating the response of the ELN/LIMS into the prompt, this limits how much context (i.e., how many experiments/samples) the system can be aware of. One potential remedy is to use retrieval-augmented generation, where the entries are embedded in a vector store and the agent will be able to query this database on put (parts of) the most relevant entries into the prompt.

### D. BOLLaMa

The field of chemistry is continuously evolving towards sustainability, with the optimization of chemical reactions being a key component [53]. The selection of optimal conditions, such as temperature, reagents, catalysts, and other additives, is challenging and time-consuming due to the vast search space and high cost of experiments [54]. Expert chemists typically rely on previous knowledge and intuition, leading to weeks or even months of experimentation [55].

Bayesian optimization (BO) has recently been applied to chemistry optimization tasks, outperforming humans in optimization speed and quality of solutions [55]. However, mainstream access to these tools remains limited due to requirements for programming knowledge and the numerous parameters these tools offer. To address this issue, we developed BOLLaMa. This artificial intelligence (AI)-powered chatbot simplifies BO for chemical reactions with an easy-to-use natural language interface, which facilitates access to a broader audience.

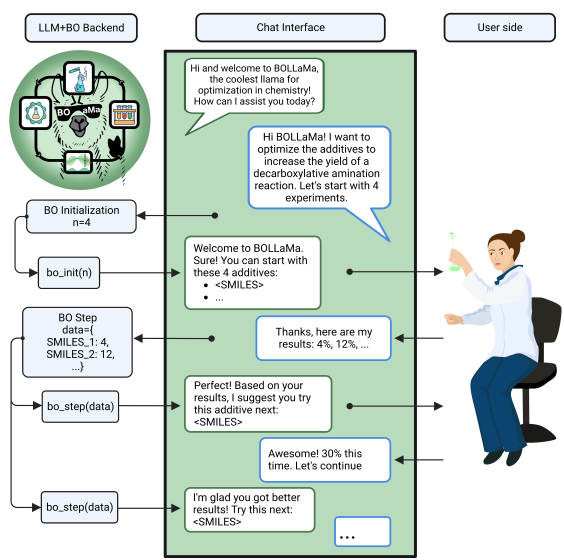

Figure 11. *Schematic overview of BOLLaMa.* A LLM can act as an interface to a BO algorithm. An experimental chemist can bootstrap an optimization and then, via a chat interface, update the state of the simulation to which the bot responds with the recommended next steps.

BOLLaMa combines LLMs with BO algorithms to assist chemical reaction optimization. The user-friendly interface allows even those with limited technical knowledge to engage with the tool. BOLLaMa's current implementation provides two main tools: the initialization function and the optimization step function [56], that are retrieved on LLM-demand as shown in Figure 11.

The primary contribution of this project is democratizing access to advanced BO techniques in chemistry, promoting widespread adoption of sustainable optimization tools, and impacting sustainability efforts within the community. This approach can be further enhanced to provide a more comprehensive assistant experience, such as with additional recommendations or safety warnings, and improve the explainability of the BO process to foster user trust and informed decision-making.

Key insights gained from this project include the critical role of accessibility in developing expert tools and the potential of LLMs in chemistry through various agent architectures [50]. In addition, the initial BO tool adapted for BOLLaMa was designed for closed-loop automated laboratories, emphasizing the need for accessible tools catering to diverse user backgrounds.

*One sentence summaries*

*a. Problem/Task* Giving scientists without coding and machine learning expertise access to Bayesian optimization.

*b. Approach*   LLM as a chat-interface for a Python package for Bayesian optimization by using ReAct-like approach in which the LLM has access to text-description of relevant functions (such as initialization and stepping of the BO run).

*c. Results and Impact*   The chat interface can successfully initialize a BO run and then convert observations reported in natural language into calls to the stepping function of the BO tool.

*d. Challenges and Future Work*   As most LLM agents, the tools suffers from robustness issues and the correct functioning cannot be guaranteed for all possible prompts.

## III. Knowledge Extraction

### A. InsightGraph



Figure 12. *The Insight Graph interface.* A suitably prompted LLM can create knowledge graph representations of scientific text that can be visualized using tools such as neo4j's visualization tools. [57]

The traditional method of performing a literature review involves months of reading relevant articles to find crucial information on material properties, structure, reaction pathways, and applications. Knowledge graphs are sources of structured information that enable data visualization, data discovery, insights, and downstream machine-learning tasks. Knowledge graphs extracted from published scientific literature covering broad materials science domains [58] as well as more-focused domains such as polymer nanocomposites [59] empower material scientists to discover new concepts and accelerate research. Until recently, capturing complex and hierarchical relationships for a knowledge graph within the materials science literature was a time-consuming effort, often spanning multi-disciplinary collaborations and many Ph.D. years. By leveraging zero to few-shot training and pre-trained LLMs, it is now possible to rapidly extract complex scientific entities with minimal technical expertise [58, 60, 61]. We envision that knowledge graphs built by LLMs based on scientific publications can offer a concise and visual means to launch a literature review.

To demonstrate a proof of concept of a zero-shot entity and relationship extraction, we identified 200 abstracts on polymer-nanocomposite materials for which detailed structured information was already available [62]. Each abstract was fed as a prompt to GPT-3.5-turbo, a language model powering the popular ChatGPT web application by OpenAI. The instructions in our prompt consisted of an example JSON containing high-level schema and information on possible entities and pairwise relationships. The nodes and relationships in the output JSON response were then stored in a neo4j graph database using Cypher, a graph query language (Figure 12). [57] The zero-shot capabilities of the model allowed the specification of an arbitrary entity and relationship types depending upon the information contained in the text. Given that this required a change in the neo4j pipeline every time the prompt changed, we found it necessary to constrain the JSON schema to a standard format.

While large language models on their own are prone to hallucinations, leveraging them with guidance to create structured databases empowers chemists/materials scientists with no expertise in natural language processing to search and build on existing knowledge leading to new insights. The speed at which LLMs can create structured graphs dramatically exceeds the years required for humans to manually curate data into existing knowledge graphs. Access to structured databases will accelerate the pace of data-driven material science research, synthesizing details embedded in dispersed scientific publications. Additionally, other scientific fields could benefit from a similar use of LLMs to extract entities and relationships to build knowledge graphs.

Owing to the non-deterministic nature of LLMs, we found that the output response would vary even when the same prompt was provided. An instruction constraining the JSON schema minimized the variability. A systematic study comparing different foundation models, prompt techniques (zero-shot, one-shot, few-shot), prompt chaining, and the role of fine-tuning is needed to evaluate the precision and recall of extracted entities

and relationships. Notably, pairwise links between the nodes are not often enough to model the complex nature of materials requiring improvement in the input schema.

*One sentence summaries*

    *a. Problem/Task*   Extraction of entities and their relationships from text.

    *b. Approach*   Prompting of GPT-3.5-turbo prompted with abstract and example JSON and the task to extract entities and their relationships in a structure as provided in the example.

    *c. Results and Impact*   The approach can successfully create meaningful JSON data structures with extracted entities and their relationships for hundreds of abstracts.

    *d. Challenges and Future Work*   The non-deterministic behavior of LLMs can lead to variability and fragile behavior. To better understand this as well as the performance of this approach, more systematic benchmarking is needed.

### B. Extracting Structured Data from Free-form Organic Synthesis Text

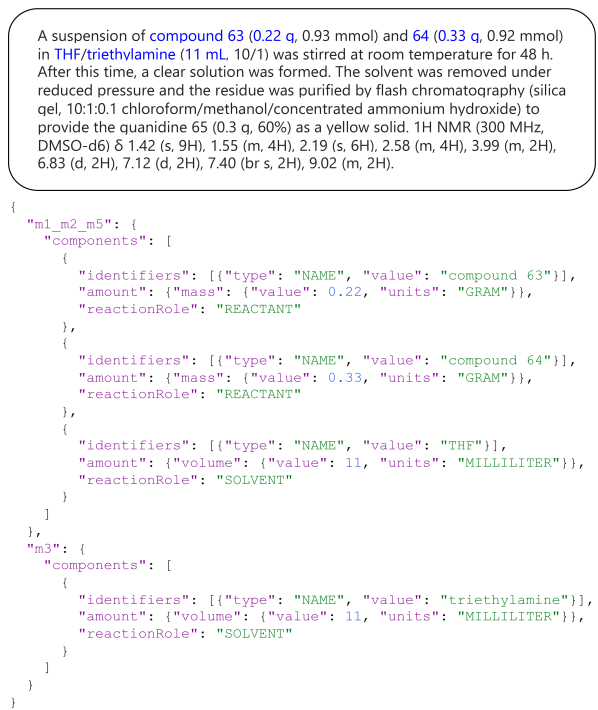

Figure 13. *The Organic Synthesis Parser interface.* The top part shows text describing an organic reaction (`https://open-reaction-database.org/client/id/ord-1f99b308e17340cb8e0e3080c270fd08`), which the finetuned LLM converts into structured JSON (bottom). A demo application can be found at `https://qai222.github.io/LLM_organic_synthesis/`.

*a. Problem*    As data-driven approaches and machine learning (ML) techniques gain traction in the field of organic chemistry and its various subfields, it is becoming clear that, as most data in chemistry is represented by unstructured text, the predictive power of these approaches is limited by the lack of structured, well-curated data. Due to the large corpus of organic chemistry literature, manual conversion from unstructured text to structured data is unrealistic, making software tools for this task necessary to improve or enable downstream applications, such as reaction prediction and condition recommendation.

*b. Solution*    In this project, we leverage the power of fine-tuned LLMs to extract reactant information from organic synthesis text to structured data. 350 reaction entries were randomly selected from the Open Reaction Database (ORD) [63]. The field of `reaction.notes.procedure_details` is used as the input (prompt), and the field of `reaction.inputs` is used as the output (completion). 300 of these prompt-completion pairs were used to fine-tune a GPT-3 (OpenAI Davinci) model using the OpenAI command line interface (version 0.27.2), and the rest were used for evaluation. In addition to this, we also explored fine-tuning the Alpaca-LoRA model [16, 64, 65] for this task. All data and scripts used in this project are available in the GitHub repository.

*c. Results and Discussion*    Surprisingly, the pre-trained language model (OpenAI Davinci), fine-tuned with only 300 prompt-completion pairs, is capable of generating valid JSON complying with the ORD data model. For the 50 prompt-completion pairs in evaluation, 93% of the components in reaction inputs were correctly extracted from the free text reaction description by the GPT-3 based model. The model also associates existing properties, such as volume or mass used in the reaction, to these components. In addition to recognizing in-text chemical entities (such as molecule names), as shown in Figure 13, tokens referencing external chemical entities (compound numbers) can also be captured by the model. On the other hand, while completing the prompts with extracted chemical information, the fine-tuned Alpaca-LoRA model was unable to properly construct a valid JSON complying with the ORD data model.

Despite these encouraging preliminary results, there are still challenges to a robust synthesis text parser. One of them is the ambiguous and often artificial boundary between descriptions of reactions and workups, which leads to misplaced chemical entities in the structured data, e.g., a solvent used in the extraction of products is instead labeled as a reaction solvent. The aforementioned external reference problem, where a compound number in the procedure is only explicitly identified in an earlier section of the manuscript, can only be solved by prompting the LLM with multiple paragraphs or even the entire document, adding more irrelevant tokens to the prompt. It is also important to prevent the LLM from "auto-completing" extracted named entities with information outside the prompt, e.g., the chemical is extracted as "sodium chloride" in the completion while it is only specified as "chloride" in the prompt.

*One sentence summaries*

*d.  Problem/Task*  Extraction of structured reaction condition and procedure data from text.

*e.  Approach*  Fine-tuning of LLMs on hundreds of prompt (unstructured text)- completion (extracted structured data) pairs.

*f.  Results and Impact*  OpenAI's davinci model can extract the relevant data with a success rate of 93 %.

*g.  Challenges and Future Work*  Parameter efficient fine-tuning could not match the performance of OpenAI's models. In addition, there are instances in which the LLM goes beyond the specified tasks (e.g., modifies/"autocompletes") extracted entries, which can lead to fragile systems.

## C. TableToJson: Extracting structured information from tables in scientific papers

Much of the scientific information published in research articles is presented in an unstructured format, primarily as free text, making it a difficult input for computational processing. However, relevant information in scientific literature is not only found in text form. Tables are commonly employed in scientific articles, e.g., to collect precursors and raw materials' characteristics, synthesis conditions, synthesized materials' properties, or chemical process results. Converting this information into a structured data format is usually a manual time-consuming and tedious task. Neural-network-based table extraction methods and optical character recognition (OCR) [66], which can convert typed, handwritten, or printed documents into machine-encoded text, can be used to extract information from tables in PDF files. However, it is often not straightforward to extract the data in the desired structured format. Nonetheless, structured data is essential for creating databases that aggregate research results, and enable data integration, comparison, and analysis.

In this context, JSON is a widely adopted structured data format due to its simplicity, flexibility and compatibility with different programming languages and systems. However, obtaining structured data following a specific JSON schema with models can be challenging. The generated JSON needs to be syntactically correct and conform to a schema that defines the JSON's structure. Models typically do not provide structured output that perfectly matches the desired JSON schema. Some manual post-processing or data transformation is often necessary to map the extracted information to the appropriate schema fields.

In this work, we have studied two approaches to generate structured JSON from data contained in tables of scientific papers focused on different research topics within the field of chemistry [67–73]. The Python `json` module was used to parse JSON data and validate the outputs.

As a first approach, the OpenAI `text-davinci-003` model was used to generate structured JSON from data in tables. The input to the LLM is the HyperText Markup Language (HTML) code of the table, obtained directly from the digital object identifier (DOI) of the article using the Python `selenium` library, while the output of the model is the data extracted in JSON form (Figure 14). The OpenAI `text-curie-001` model, although not tested in this work, can also be utilized if the number of input tokens, considering both the HTML text of the table and the schema, meets the requirements of this model (máximum 2049 input tokens, compared to 4097 for `text-davinci-003`).

The use of the OpenAI model to generate structured JSON was compared with a second approach, i.e., the use of `jsonformer` (https://github.com/1rgs/jsonformer), which implements a data processing pipeline that combines the model generation with appropriate data transformation. This method introduces an efficient way to generate structured JSON using LLMs by generating only the content tokens and filling in the fixed tokens. This avoids generating a complete JSON string and parsing it. This approach ensures that the produced JSON is always syntactically correct and aligns with the specified schema. [74]

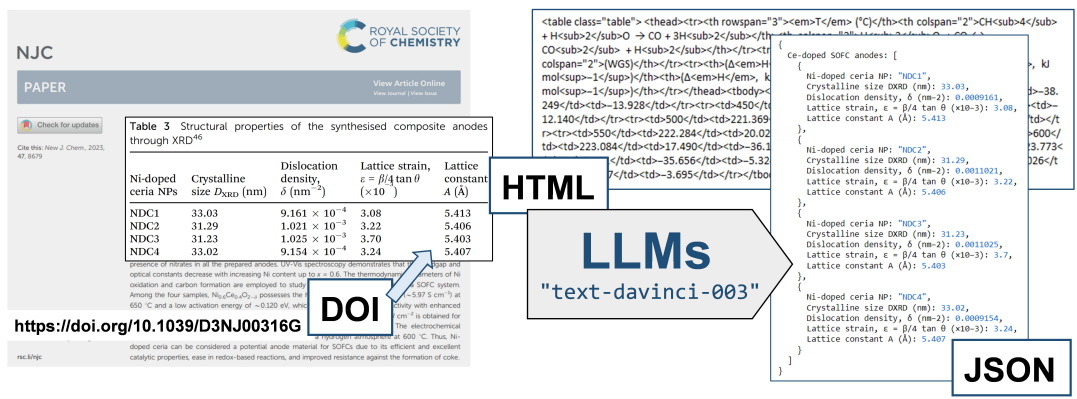

Figure 14. *TableToJson*. Extraction of structured information from scientific data in tables using LLMs. The input to the LLM model is the HTML code of a table contained in a scientific paper. The output of the LLM model is data structured in JSON form. Results can be visualized in this demo app: https://vgvinter-tabletojson-app-kt5aiv.streamlit.app/.

In our first approach, we directly asked the OpenAI `text-davinci-003` model to generate a JSON object according to a desired JSON schema provided in the model prompt. The table content was also included in

the prompt as HTML code. The accuracy in the prediction, calculated as the percentage of schema values generated correctly, is shown in Figure 15. In all examples, the OpenAI model was queried with a simple prompt, and it correctly extracted all the data in the table and inserted every value into the corresponding position in the schema, with 100 % accuracy, providing as output a JSON object. This model also correctly generated both string and number values according to the type assigned in the schema. However, in two of the examples, the OpenAI model did not generate the JSON object name specified in the schema when the corresponding name was not found in the table, generating only the list of components. This was solved by modifying the object name in the schema to a term that more closely aligned with the content of the table. It appears that when the model could not establish a clear relationship between the provided name and the table content, it disregards that part of the schema during generation. These results indicate that the OpenAI `text-davinci-003` model is able to convert scientific data from tables of research papers to a structured format following the approach used in this work, where the desired JSON schema was included in the model prompt. Nevertheless, the model retains a certain degree of freedom to modify the requested scheme if it considers that something may be wrong.



Figure 15. *TableToJson*. Results of the structured JSON generation of tables contained in scientific articles. Two approaches are compared: (i) the use of an OpenAI model prompted with the desired JSON schema, and (ii) the use of an OpenAI model together with **jsonformer**.

The second approach used to generate structured information was a version of the `jsonformer` approach adapted for use with OpenAI LLMs (`https://github.com/martinezpl/jsonformer/tree/add-openai`), with the implementation of the inclusion of the table text as an input parameter to the `jsonformer` function.

Detection of strings indicating null values was also added when the schema type is number, as "nan", "NaN", "NA", and "NAN" entries are common in research data tables. The OpenAI `text-davinci-003` model was used. In this case, the model was prompted with the desired JSON schema and the HTML code of the studied table. `Jsonformer` reads the keys from the JSON schema and only delegates the generation of the value tokens to the language model, ensuring that a valid JSON is generated by the LLM model.

For this approach, the accuracy in the prediction is also shown in Figure 15. The use of the OpenAI `text-davinci-003` model together with `jsonformer` generated valid JSON objects with 100% accuracy for most of the tables evaluated using a simple prompt. Figure 16 shows the results of one of the examples studied, where using a simple descriptive prompt denoting the type of input text, this approach correctly generated structured data JSON from a table with a complex header. However, it was detected that when the values to be generated contain special characters or specific texts, a more detailed prompt with some simple examples, but without finetuning, can be necessary to provide good results, as shown in Figure 17 for a special numeric notation that included power numbers.



Figure 16. *TableToJson*. Structured JSON generation of tables contained in scientific articles using a prompt with a simple description of the type of input text. One example is shown for a table that contains data on properties of biomass materials [71].

As shown in Figure 15, in one of these examples, an accuracy of 94% was obtained from a table containing a few catalyst names that included the "-" character, and those values were erroneously generated. In another example, an accuracy of 80% was initially obtained due to errors in the generation of numbers with powers (e.g., $9.161 \times 10^4$), which could be solved by adding an explanation in the prompt: `"if you find numbers as 1.025 × 10<sup>3</sup>, this means 1.025e-3"`, increasing the accuracy to 100%.

Next, a table with more complex content (long molecule names, hyphens, power numbers, subscripts, and superscripts...) was selected (Figure 15), resulting in an accuracy of 46% in the JSON generation, meaning that only 46% of the schema values were correctly generated. The erroneous generation of long formula or molecule names with a mixture of letters and numbers as subscripts could be solved by increasing the value of the `max_string_token_length` argument of the `jsonformer` function to get a longer response where the end of the string can be detected more easily, which increased the accuracy to 60%. `Jsonformer` also showed some issues in this example in generating power numbers, which are represented as `10<sup>−n</sup>` in the input HTML text. As mentioned above, this was solved by adding a specific explanation in the prompt, increasing the accuracy to 86%. A specific explanation was also included in the prompt to address the issues related to the presence of hyphens in the text. Still, this problem could not be solved systematically, and the resulting accuracy varied between 86% and 100% for several JSON generation attempts. In this particular case, the generated value provided by the model included Unicode text instead of the "-" character (and usually several "\" characters). An instruction to `"decode Unicode characters in your response"`

was then included in the prompt. Although this solution sometimes yielded satisfactory results, it did not systematically guarantee correct output. These results indicate that the OpenAI model combined with `jsonformer` can provide wrong outputs when the values to be generated contain some special characters, such as the "-" character in this example. This issue requires further investigation to be improved.



Figure 17. *TableToJson*. Structured JSON generation of a table contained in a scientific article using a standard prompt and a prompt with a few simple examples of the special numeric notation found in some of the cells of the input table [72].

Lastly, for one of the examples, a test was performed by providing a wrong schema to the model (Figure 15). In this case, as expected, `jsonformer` inserted the values contained in the table into the given wrong schema in a more or less ordered fashion, generating an invalid output. However, the OpenAI model created a new schema according to the table structure and headers, providing a valid result, and confirming its freedom to decide what may be wrong with the user's query. An example of these results is shown in Figure 18.

The two approaches used in this work showed a good performance in the generation of JSON format when the data contained in the table are regular strings or numbers, with an accuracy of $100\%$ in most of the examples. The results of this work show that, although the OpenAI `text-davinci-003` is able to easily extract structured information from tables and give a valid JSON output, this approach cannot guarantee that the outputs will always follow a specific schema. On the other hand, although `jsonformer` may present problems when special characters need to be generated, some of these issues have been solved with careful prompting, and others could probably be solved with further research. It can be concluded that `jsonformer` can be a powerful tool for the generation of structured data from unstructured information in most tables, ensuring the generation of valid JSON syntax as the output of LLMs that always complies with the provided schema. The use of `jsonformer` could facilitate and promote the creation of databases and datasets for numerous topics within the field of chemistry, especially in experimental domains, where the availability of structured data is very scarce.

*One sentence summaries*

a. *Problem/Task*  Extracting structured data in a JSON-schema-compliant form from HTML tables.

Figure 18. *TableToJson.* Results of the structured JSON generation of a table after providing the model with a wrong schema. The output generated using the OpenAI model together with `jsonformer` is shown on the left (values in the table are inserted in the provided wrong schema), while the output generated using directly the OpenAI model is shown on the right (a new schema is created following the table content).

*b. Approach*   Two approaches were compared: Direct prompting of OpenAI's `text-davinci-003` model with the input table and the JSON schema, as well as the `Jsonformer` approach, which only samples from a subset of tokens in field-wise generation steps.

*c. Results and Impact*   Both approaches can extract data in schema-compliant from tables with high success rates. Due to hard-coded decoding rules, `Jsonformer` failed in some cases.

*d. Challenges and Future Work*   While the `Jsonformer` approach can guarantee valid syntax, it can fail in cases that were not considered in the development of the decoding rules. Hence, future work is needed for increasing the general applicability of constrained decoding strategies.

### D. AbstractToTitle & TitleToAbstract: text summarization and text generation

#### 1. Problem

Text summarization and text generation are some of most the common tasks in natural language processing (NLP). Often it is tricky to obtain well-defined and curated datasets for these tasks. Also, evaluating the performance of an NLP model is challenging because there is no unique way to summarize and generate text. Luckily, there are many publicly available manuscripts for chemistry and materials science in open access platforms such as arXiv and PubChem. These datasets can be used along with LLMs to solve problems such as: 1) given title of the article; generate an abstract, 2) given an abstract; generate a title, which should be similar to the one available in the literature. Such models can greatly help in technical writing, especially with a human expert in the loop.

#### 2. Solution

The above challenging tasks require an organized infrastructure for curated data and tools. JARVIS-ChemNLP [75] is an integrated library for materials chemistry text data for NLP tasks such as text classification, token classification, abstractive summarization, text generation, and integrating with existing DFT databases. ChemNLP uses transformer models to generate titles from abstracts and vice versa for arXiv condensed matter category articles.

There is no unique approach for both tasks. Nevertheless, highly qualified scientists and engineers generate such abstracts and titles and can act as a valuable benchmark to evaluate model performance. Moreover, developing metrics for text generation is a difficult task. Here, for the sake of simplicity, we choose the Recall-Oriented Understudy for Gisting Evaluation (ROUGE) metrics for evaluating the model performance.

ChemNLP uses models such as a pre-trained Text-to-Text Transfer Transformer (T5) developed by Google for text summarization. The pre-trained model is further fine-tuned to produce summaries of abstracts and analyze the performance by analyzing its closeness to the title of the given paper.

For the text generation, ChemNLP fine-tunes LLMs such as the pre-trained GPT-2-medium model from the HuggingFace library.

For these tasks, the community can contribute their models on a benchmarking platform called the JARVIS-Leaderboard (`https://pages.nist.gov/jarvis_leaderboard/`). Specifically, the TextGen task is available at: `https://pages.nist.gov/jarvis_leaderboard/AI/TextSummary/arxiv_summary_text/` and TextSummary task is available at `https://pages.nist.gov/jarvis_leaderboard/AI/TextGen/arxiv_gen_text/`

#### 3. Example

Title to Abstract example for `cond-mat/0105402`. [76]

| Question |
| --- |

MgB2 under pressure: phonon calculations, Raman spectroscopy, and optical reflectance can be described as

| Fine-tuned GPT-2-medium model output |
| --- |

The amplitude of phonon vibrations in a material exhibiting an unconventional lattice structure has been measured in the low pressure region of the granular MgB2 phase. The peak in the high pressure region is the first measurement of phonon amplitude ever presented in a granular superconducting material, and it indicates a potential source of superconductivity. Superconducting Transition from Fluctuations to Brillouin T-Wings in CeFo Josephson Junctions can be described as We report electronic and numerical calculations for solving the electron-wave (E-W) conduction spectra of a quasi-two-dimensional superconducting monolith CeFo Josephson junctions: The amplitude of

phonon vibrations in the low pressure region (KPSM), and in the high pressure region of the superconducting density wave (SDW) densities, are used to understand the superconductivity.

**Actual abstract**

The effect of pressure on optical phonon frequencies of $MgB_2$ has been calculated using the frozen-phonon approach based on a pseudopotential method. Gruneisen parameters of the harmonic mode frequencies are reported for the high-frequency zone-center $E_{2g}$ and $B_{1g}$ and the zone-boundary $E_{2u}$ and $B_{2u}$ modes at $A$. Anharmonic effects of phonon frequencies and the implications of the calculated phonon frequency shifts for the pressure dependence of the superconducting transition temperature of $MgB_2$ are discussed. Also reported are Raman and optical reflectance spectra of $MgB_2$ measured at high pressures. The experimental observations in combination with calculated results indicate that broad spectral features we observed in the Raman spectra at frequencies between 500 and 900 $cm^{-1}$ cannot be attributed to first-order scattering by zone-center modes, but originate in part from a chemical species other than $MgB_2$ at the sample surface and in part from a maximum in the $MgB_2$ phonon density of states. Low-temperature Raman spectra taken at ambient pressure showed increased scattering intensity in the region below 300 $cm^{-1}$.

Interestingly, the generated abstract contains grammatically and syntactically incorrect sentences. We suspect that this is due to our use of a small, outdated, base model. However, more systematic analysis will need to be performed in future work.

*One sentence summaries*

*a. Problem/Task*  Text summarization and generation, in specific, a summary of an abstract into a title and generation of an abstract conditioned on a title.

*b. Approach*  Fine-tuning of transformer models such as T-5 and GPT-2 on data from arXiv.

*c. Results and Impact*  Initial exploration indicates that transformer models might be suitable for this task.

*d. Challenges and Future Work*  More systematic analysis, including rating of the generated titles and abstracts by domain experts is required to identify the limitations of this approach.

### IV. Education

#### A. i-Digest

*a. Problem* Over the last few years, especially during the Covid period, most of us had to switch to the online mode of working in our day-to-day jobs. And even today, the online mode of working has, to some extent, stayed on as it turned out to be convenient for both employers and employees. One clear example can be found in the field of education, where the use of video lectures became the norm for teaching students in universities and schools. Likewise, podcasts and three-minute thesis videos, which communicate important scientific information to society at large, have grown tremendously [77, 78]. This has led to a situation where, at present, we have an enormous amount of important scientific information stored in the form of videos and audio all over the internet. A current challenge is to summarize and make use of this knowledge efficiently. Some efforts in this direction have been made by using AI Youtube summarizers and QnA Bots [79]. We would like to build upon such efforts and create a tool for the field of education.

*b. Solution* We present a tool that self-guides students and other users toward a better understanding of the content of a video lecture or a podcast. In order to accomplish this, we used publicly available LLMs like Open AI's Whisper [80] and GPT-3.5-turbo model. All the user needs to do is provide a link to the lecture video or audio file. After only a short time, the overview page shows some technical keywords on which the video is based, a short but comprehensive summary, and some questions for the user to assess his or her understanding of the concepts discussed in the video/audio (Figure 19). Additionally, for chemistry enthusiasts, if some chemical elements/molecules are discussed in the content, we link them to online databases. At the backend, we first convert the video to audio using Pytube (In the case of a podcast, this step is not needed). Then we use the Whisper model to transcribe the audio to text. Next, we make use of the OpenAI GPT-3.5-turbo model to obtain a short summary and a set of questions based on the text. Finally, we extract the name of chemical elements/molecules and list the PubChem database entry for that element/molecule on the overview page. [81–83] The web interface was made using the open-source app framework Streamlit [84].

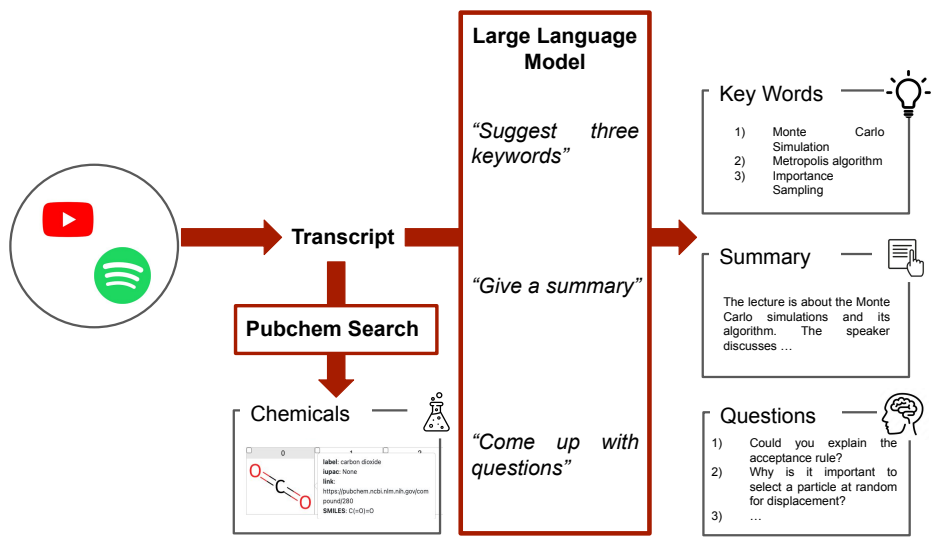

Figure 19. A schematic of the i-digest interface. On providing a link to an online video or audio, i-digest generates some technical keywords, a short but comprehensive summary, and a list of questions based on the content in the video/audio. Additionally, chemicals discussed in the content are linked to online databases such as PubChem.

*c. Impact* We strongly believe that extracting important scientific information in terms of short lecture notes and questions would help to push forward the field of education towards creating and using resources more efficiently. Moreover, by providing additional links to resources, e.g., databases, journals, and books,

we provide an opportunity for the user to go beyond the content of the lecture and spark interest in a more detailed understanding of the topic. Specifically, this would help researchers/teachers/professors to create new course content or to update/modify already available content. In general, our tool covers a broad range of users, from the youngest learner to the chemistry novice who wants to kickstart his research, all the way to professors, course creators, and lifetime learners.

*d. Lessons learned* Working together with colleagues can be fun and enriching and often help to solve big problems. This hackathon taught us that even in one day, coming together can help achieve something significant.

### One sentence summaries

*e. Problem/Task* Provide students with automatically generated active learning tasks for lecture recordings.

*f. Approach* Transcription of videos using OpenAI's Whisper model, prompting of OpenAI's GPT-3.5-turbo model to produce a short summary and questions based on the transcript, as well as to extract mentions of chemicals in the text.

*g. Results and Impact* The system can transcribe the text, generate meaningful questions, and successfully extract mentions of chemicals.

*h. Challenges and Future Work* It is difficult to systematically evaluate the performance of this system due to the lack of suitable benchmarks/eval. An obvious extension of this approach is to condition it on further material (e.g., lecture notes and books). In addition, one might automatically score the answers and show them at the beginning and at the end of the video. This would allow us to evaluate the learning of the students and to guide them to the relevant material in case a question was not answered correctly.

## V.   Meta analysis of the workshop contributions

We have a female/male ratio of about 30 % among the workshop participants who co-authored this paper. We have participants from 22 different institutions in 8 countries.

Most teams combine expertise from different institutions (Figure 21), in several cases beyond academia (Figure 22). Around 20 % of the teams are international, with participants from two countries (Figure 23).

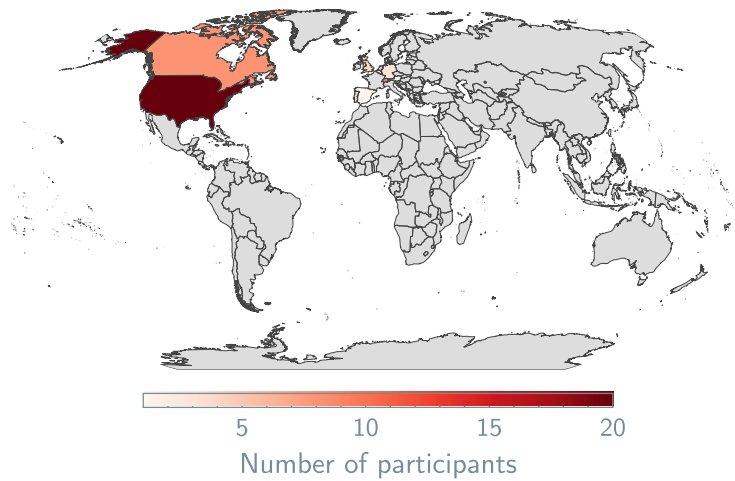

Figure 20. Worldmap (Robin projection) with the number of participants shown in color.

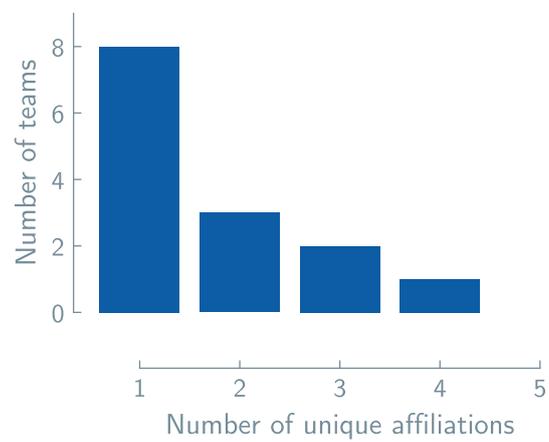

Figure 21. Histogram of the number of unique affiliations per team.

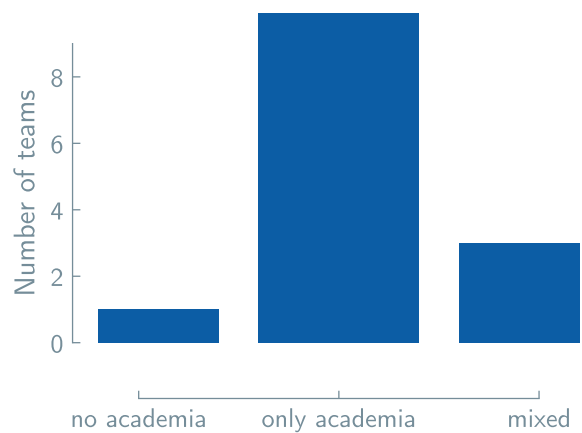

Figure 22. Number of teams with participants only from academia or academia and industry/nonprofit, respectively. We counted national labs as "academia".

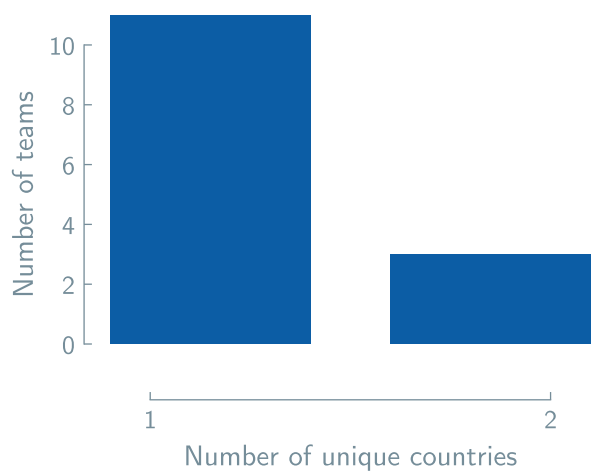

Figure 23. Histogram of the number of unique countries per team.

**Acronyms**

**AI:** artificial intelligence.

**API:** application programming interface.

**BO:** Bayesian optimization.

**CAS:** Chemical Abstract Services.

**COT:** chain of thought.

**DFT:** density functional theory.

**DOI:** digital object identifier.

**ELN:** electronic lab notebook.

**GA:** genetic algorithm.

**GPR:** Gaussian process regression.

**GPT:** generative pretrained transformer.

**GUI:** graphical user interface.

**HTML:** HyperText Markup Language.

**ICL:** in-context learning.

**ID:** inverse design.

**InChI:** international chemical identifier.

**JSON:** JavaScript object notation.

**LIFT:** language-interfaced fine-tuning.

**LIMS:** laboratory information system.

**LLM:** large language model.

**LoRA:** low-rank adaptors.

**MAD:** median absolute deviation.

**MAE:** mean absolute error.

**MAPI:** Materials Project API.

**ML:** machine learning.

**NER:** named entity recognition.

**NLM:** national library of medicine.

**NLP:** natural language processing.

**OCR:** optical character recognition.

**ORD:** Open Reaction Database.

**PDB:** protein data bank.

**PEFT:** parameter efficient fine-tuning.

**RF:** random forest.

**RLHF:** reinforcement learning from human feedback.

**ROUGE:** Recall-Oriented Understudy for Gisting Evaluation.

**SELFIES:** self-referencing embedded strings.

**SMILES:** simplified molecular-input line-entry system.

**SVM:** support vector machine.

**UI:** user interface.



\end{document}